\documentclass{aa}
\usepackage{graphicx}
\usepackage[varg]{txfonts}

\usepackage{latexsym}
\usepackage{natbib}
\usepackage{amssymb}
\usepackage{amsmath}

\newcommand{\re}{$R_e$}
\newcommand{\Lsig}{$\log(L)-\log(\sigma)$}
\newcommand{\Lsigb}{$L=L'_0\sigma^{\beta}$}

\newcommand{\MR}{$\log(R_e)-\log(M^*)$}

\newcommand{\muem}{$\langle\mu\rangle_e$}
\newcommand{\Ie}{$\langle I\rangle_e$}

\newcommand{\IeRe}{$\log(\langle I\rangle_e)-\log(R_e)$}
\newcommand{\muere}{$\langle \mu\rangle_e-\log(R_e)$}
\newcommand{\FPR}{$\log(\sigma)-\log(\langle I\rangle_e)-\log(R_e)$}

\newcommand{\kfilt}{$\rm{K}$}
\newcommand{\jfilt}{$\rm{J}$}
\newcommand{\bfilt}{$\rm{B}$}
\newcommand{\vfilt}{$\rm{V}$}

\newcommand{\Sers}{$r^{1/n}$}

\newcommand{\ie}{{\em i.e.}}
\newcommand{\eg}{{\em e.g.}}

\newcommand{\muerespace}{{$\log(\langle I\rangle_e) - \log(R_e) - \log(\sigma)$}}

\begin{document}

	\title{The parallelism between galaxy clusters and early-type galaxies:}
	\subtitle{II. Clues on the origin of the scaling relations}




   \author{M. D'Onofrio \inst{1,2} \and
	C. Chiosi \inst{1} \and
	M. Sciarratta \inst{1} \and
	P. Marziani \inst{2}
}

\institute{Department of Physics and Astronomy, University of Padua,
	Vicolo Osservatorio 3, I-35122 Padua, Italy \\
	\email{mauro.donofrio@unipd.it}
	\and
	INAF -- Osservatorio Astronomico di Padova, Vicolo Osservatorio 5, I-35122 Padova, Italy \\
}

\date{Received April, 2020; Accepted xx}

\abstract
{This is the second work dedicated to the observed parallelism between galaxy clusters and early-type galaxies.
{The focus is on the distribution of these systems in the scaling relations (SRs) observed when effective radii,
effective surface brightness, total luminosities and velocity dispersions are mutually correlated.}}
{{Using the data of the Illustris simulation we try to speculate on the origin of the observed SRs.}}
{{We compare the observational SRs extracted from the database of the WIde-field Nearby Galaxy-cluster Survey (WINGS)
with the relevant parameters coming from the Illustris simulations. Then we use the simulated data at different redshift to
infer the evolution of the SRs.}}
{The comparison demonstrate that galaxy clusters (GCs) at $z\sim0$ follow the same \Lsig\ relation of early-type galaxies (ETGs)
and that both in the \IeRe\ and \MR\ planes the distribution of GCs is along the sequence defined by the brightest and massive
early-type galaxies (BCGs). The Illustris simulation reproduces the tails of the massive galaxies visible both in the
\IeRe\ and \MR\ planes, but fail to give the correct estimate of the effective radii of the dwarf galaxies that appear too l
arge and those of GCs that are too small. The evolution of the SRs up to $z=4$ permits to reveal the complex evolutionary
paths of galaxies in the SRs and indicate that the line marking the Zone of Exclusion (ZoE), visible both in the \IeRe\ and
\MR\ planes, is the trend followed by virialized and passively evolving systems.}
{{We speculate that the observed SRs originate from the intersection of the virial theorem and a relation \Lsigb\ where
the luminosities depend on the star formation history. }}

\keywords{Galaxy clusters --
	 Early-type galaxies --
	 Galaxy structure --
	 Galaxy photometry --
	 Galaxy scaling relations --
	  Numerical simulations
}

\maketitle

\section{Introduction} \label{sec:intro}

\noindent
The scaling relations (hereafter SRs), \ie\ the 2D/3D correlations among the
parameters describing the stellar systems, are very important tools to understand their formation and evolution.
These relations do not enter in any physical theoretical model or
numerical simulation, but are used only a posteriori to test the goodness of models by means of checks between predictions
and observations.

The SRs {of galaxies} are quite easily derived from observations, but unfortunately not yet fully understood.
The most famous examples are  \eg\ the Fundamental Plane relation \FPR\ (hereafter FP,
\citealp{Djorgovski&Davis1987,Dressleretal1987}), the Faber-Jackson relation \Lsig\ (hereafter FJ,
\citealp{FaberJackson}), the Tully-Fisher relation (hereafter TF, \citealp{TullyFisher}), the surface
brightness-radius relation (hereafter \IeRe, \citealp[see e.g.][]{Kormendy1977,Donofrio2017}), the Radius-Mass relation
(hereafter \MR\ relation) \citep{ChiosiCarraro2002,Graham2013} and in general the correlations involving the color,
metallicity, shape, angular momentum, star formation rate (SFR) and the initial mass function of galaxies
\citep{DekelBirnboim2006,Duttonetal2011,Cappellari2013a,Cappellari2013b,FallRomanowsky}.

In some cases the SRs did successfully constrain models, for example in the mass-metallicity relation
\citep[see e.g.][]{Faber1973} that is strongly linked with the path of chemical evolution and with the inflow
and outflow processes, and in the black-hole (BH) - bulge-mass relation \citep[see e.g.][]{Magorrian1998},
that suggests a co-evolution of these structures.

SRs are therefore a valuable tool of investigation, even if they
represent only a snapshot of the physical properties of galaxies at the present epoch, not distinguishing
between cause and effect, past and future \citep[see e.g.,][]{Lagos2016,DFB2019}. Unfortunately the data
available for galaxies at high redshift are still sparse and not homogeneous, so that we know only
approximately the evolution of the SRs. However, thanks to the modern numerical simulations, we are
now able to predict the structure of galaxies across time and consequently the trends of the SRs at different
cosmic epochs.

Recently \cite{Cariddietal2018}, {following the historical analyses of \cite{Schaeffer} and \cite{Adami}, added an
important element to the debate on the SRs. They confirmed that galaxy
clusters follow the same distribution of early-type galaxies (ETGs) in the \IeRe\ and \Lsig\ FJ planes and
have a similar color-magnitude diagram. A similar result was obtained by \cite{Donofrio2013}, who
noted that clusters and ETGs share the same FP relation.}
This means that on different scales the processes shaping the properties of galaxies and clusters are quite similar.

Moved by this intriguing observational evidence \cite{Donofrio2019} (hereafter paper I) started a detailed analysis of
the parallelism between these systems. They showed that GCs and ETGs share a similar behavior
in the luminosity/mass growth curves and in the surface brightness/mass profiles,
once these are normalized to the effective radius enclosing half the total luminosity and half-mass radius
respectively. The profiles can be easily superposed with a small scatter.
The S\'ersic's law \Sers\ fits very well the bulk of the luminosity and mass distribution of ETGs and
clusters, but fails in the inner and outer regions, where numerous physical effects are at work. In the
center, feedback effects from supernov\ae\ (SNe) and active galactic nuclei (AGN) can significantly change
the luminosity distribution, while in the outer regions mergers can alter the shape of the profiles. The mass profiles
are also affected by the presence of the baryon component in the same regions. The range of values of the S\'ersic index
$n$ is quite large both in ETGs and clusters. For ETGs $n$ increases systematically from faint and low mass objects to
bright and massive ones, while for clusters this trend is less evident.

These striking parallelisms between systems so different in size (from the kpc to the Mpc scale) is far
from being fully understood, considering the different processes that might drive the evolution of galaxies
and clusters in the SRs. In this framework it is therefore important to inspect in a more detailed way the main SRs shared
by these systems.
This analysis might have a relevant cosmological impact, in particular for
understanding the relative contribution of dark and luminous matter in the formation and evolution of these
structures. One can in fact address the relative importance of dissipational and dissipationless merging processes, the role
of mass stripping and that played by star formation and feedback effects.

The aim of this paper is to provide a qualitative comparison of the behaviour of galaxies and GCs in
the main SRs. We will discuss in particular the parallelism observed between clusters and ETGs, showing that
these systems share a similar distribution in the SRs. We will also show that the data of the Illustris
numerical simulation \citep{Vogel2014} reproduce the main features of the SRs of galaxies and  give important
insights on the evolution of the SRs at different cosmic epochs. We decided for this approach because the
Illustris simulation tracked successfully the small-scale evolution of gas and stars, reproducing the metal
and hydrogen content of galaxies, yielding for the first time a reasonable morphological mix of thousands
of galaxies. The virtual universe resembles closely the real one and can then be used to infer the mass
assembly history of galaxies and clusters.

The paper is designed as follows: in Sec. \ref{sec:1} we introduce the observed galaxy and cluster samples,
we describe the data of the Illustris simulation \citep{Vogel2014} used in this work and we clarify the use made of galaxy
luminosities and pass-bands; in Sec. \ref{sec:2} we provide a theoretical introduction
necessary to interpret the origin of the observed SRs;
in Sec. \ref{sec:3} we start the discussion of the SRs showing how they are mutually linked each other.
We describe the distribution of galaxies and clusters in the  SR planes and we address the problem of the
observed Zone of Exclusion (ZoE);
in Sec. \ref{sec:4} we exploit the numerical simulation to follow the progenitors of present day galaxies
along their evolution in the SRs; finally, conclusions are drawn in Sec. \ref{sec:5}.

Throughout the paper we assumed the standard values of the $\Lambda$-CDM cosmology \citep{Hinshaw_etal_2013}
in all our calculations:
$\Omega_m = 0.2726, \Omega_{\Lambda}= 0.7274, \Omega_b = 0.0456, \sigma_8 = 0.809, n_s = 0.963, H_0 = 70.4\,
km\, s^{-1}\, Mpc^{-1}$.

\section{The sample} \label{sec:1}
\subsection{The database of real galaxies} \label{sec:1:1}
The observational data for galaxies and clusters are those extracted from the WINGS and Omega-WINGS database
\citep{Fasano2006,Varela2009,Cava2009,Moretti2014,Donofrio2014,Gullieuszik2015,Morettietal2017,
Cariddietal2018,Bivianoetal2017}.

The WINGS and Omega-WINGS surveys are the largest and more complete data sample for galaxies in nearby clusters
($0<z<0.07$). The core of the surveys is the dataset of optical \bfilt\ and \vfilt\ images of 76 clusters,
obtained with the Wide Field Camera (WFC, $34'\times 34'$) of the INT-2.5 m telescope in La Palma (Canary Islands, Spain)
and with the Wide Field Imager (WFI, $34'\times 33'$) of the MPG/ESO-2.2 m telescope in La Silla (Chile).

The WINGS optical photometric catalog is 90\% complete at \vfilt\ $\sim 21.7$ \citep{Varela2009}. The
database includes respectively 393013 galaxies in the \vfilt\ band and 391983 in the \bfilt\ band. The cluster
outskirts were mapped with the Omega-WINGS photometric survey at the VST telescope \citep{Gullieuszik2015}
covering 57 out of 76 clusters.

The near-infrared extension of the survey WINGS-NIR \citep{Valentinuzzi2009} consists of \jfilt\ and \kfilt\
images of a subsample of 28 clusters, taken with the WFCAM camera mounted at the UKIRT telescope. Each mosaic is
$\approx\!0.79\;\rm{deg}^2$. The 90\% detection rate limit for galaxies is reached at $J = 20.5$ and $K = 19.4$.
We used these data to get the galaxy stellar masses of our galaxies using the $K$ band luminosity as a proxy.

The WINGS and Omega-WINGS surveys have got two spectroscopic follow-up:
the first includes a subsample of 48 clusters (26 in the north and 22 in the south hemisphere) done with the
spectrographs WYFFOS@WHT ($\lambda$range $= 3800\div7000$ \AA, resolution   FWHM $=3$ \AA) and 2dF@AAT
($\lambda$range $= 3600\div8000$ \AA, resolution  FWHM $=6$ \AA).
{The second is an amplification of the south sample obtained with the AAOmega spectrograph at the Australian
Astronomical Observatory (AAT) that has a resolution R$=$1300 (FWHM $=3.5\div6$ \AA) in the wavelength range is
$3800\div9000$ \AA\ \citep{Morettietal2017}. With  the spectroscopic sample we got the redshift measurements for
thousand of galaxies \citep{Cava2009,Morettietal2017}. The spectroscopic sample is 80\% complete down to V$=$20.
In this paper we used the subsample analyzed with spectro-photometric techniques to derive the SFR at different epochs,
the stellar masses $M^*$ and age, the internal extinction $A_V$ and the equivalent widths of the absorption features
\citep[see][]{Fritzetal2011}}.

The main WINGS data used here are the same of paper I. In this case we present the distribution in the SRs
for the brightest (BCG) and second brightest (II-BCG) galaxies of the clusters and for a number of faint ETGs
(DGs) belonging to the clusters that were randomly chosen in the CCD images and re-analyzed (see for details paper I).

In addition we have used several data extracted from the WINGS database \citep{Moretti2014}: \\
1. The velocity dispersions of 1729 ETGs, measured by the Sloan Digital Sky Survey (SDSS) and by the National Optical
Astronomical Observatory (NOAO) survey, already used by \cite{Donofrio2008} to infer the properties of the FP
(see that paper for all details); \\
2. The effective radii and surface brightness of 34982 galaxies, either ETGs and late-type galaxies (LTGs),
members and non members of our clusters, derived by
\cite{Donofrio2014} through the software GASPHOT \citep{Pignatelli}; \\
3. The stellar masses obtained by the
fits of the spectral energy distributions (SED) by \cite{Fritzetal2007,Fritzetal2011} or by the K-band
luminosity \citep{Valentinuzzi2009}; \\
4. The luminosity distance derived from the redshifts measured by
\citep{Cava2009,Morettietal2017}.

The corresponding parameters for the galaxy clusters are those measured by
\cite{Bivianoetal2017} and \cite{Cariddietal2018}. {The effective radii were obtained by constructing the luminosity
growth curves of the clusters starting from the central BCG, subtracting in a statistical way the background of galaxies
not belonging to the cluster. The central velocity dispersions were instead derived from the available redshifts. For all
details we refer to the above mentioned papers.} As explained in paper I we have used only the clusters with the light profiles
well fitted by the \Sers\ law for our comparison with the ETGs. We believe that the clusters with anomalous light profiles are
still suffering the consequences of recent merging events that have affected their light distribution.

In some plots we have adopted a subset of the WINGS galaxies for which the morphology and the membership were determined by
\cite{Fasanoetal2012} and \cite{Cava2009} respectively, and a small sample of faint DGs  with new measured velocity dispersions
derived by \cite{Bettoni2016}. To avoid confusion we provide in each figure a caption with the description of the WINGS galaxy
sample used.

\subsection{The database of simulated galaxies}
\label{sec:1:2}
The simulated data are those provided by the Illustris
simulation\footnote{http://www.illustris-project.org/data/}
\citep[][to whom we refer for all details]{Vogel2014,Genel_etal_2014,Nelsonetal2015}. In paper I we provided a
full description of the data extracted from the Illustris database.
We have used the run with full-physics (with both baryonic and dark matter) having the highest degree of
resolution, i.e. Illustris-1 \citep[see Table 1 of][]{Vogel2014} extracting in particular the \vfilt-band
photometry, the mass and half-mass radii of stellar particles (i.e. the integrated stellar populations), as
well as the comoving coordinates $(x',y',z')$\footnote{The more recent data of Illustris-TNG have not be used,
because they were released when this work was almost completed.}

In paper I we analyzed the projected light and mass profiles using the $z'=0$ plane as reference plane and we
adopted the non-parametric morphology of \cite{Snyder_etal_2015}. Starting from the \vfilt\ magnitudes and
positions of the stellar particles, we computed the effective radius \re\ and effective surface brightness
\muem, the radial surface brightness profile in units of $r/R_e$, the best-fit S\'ersic index and the
line-of-sight velocity dispersion $\sigma$ for BCGs, II-BCGs and random ETGs following
\citet{Zahid_etal_2018}. For GCs, we simply used the relation $\sigma^2\simeq
2GM_{200,crit}R_{200,crit}^{-1}$, where $M_{200,crit}$ and $R_{200,crit}$ are tabulated values related to the
volume enclosing 200 times the critical density of the Universe. {The data of the simulation does not permit to derive
in an easy way the central velocity dispersion of GCs.}

Furthermore, in order to follow the evolution of the SRs, we extracted from the Illustris database the
stellar mass, the \vfilt\ luminosity, the half-mass radius, the velocity dispersion and the SFR for the whole
set of galaxies (with mass $\log(M^*)\geq9$ at $z=0$) in the selected clusters at redshift
$z=0$, $z=0.2$, $z=1$, $z=1.6$, $z=2.2$, $z=3$ and $z=4$. With these data we can follow the progenitors of each object across
the epochs and compare observations with simulations up to redshift $z=4$.

\subsection{Luminosities, magnitudes, and colours}

The WINGS data for galaxies and GCs have been taken in the B and V pass-bands of the
Johnson photometric system and whenever necessary corrected for the cosmic K-corrections. They are also
reduced to  the co-moving rest-frame of the galaxies when magnitudes are translated to
absolute luminosities. Therefore speaking of observational luminosities we always refer to these pass-bands.

The theoretical simulations of the Illustris library are also given in these pass-bands so they are
homogeneous with the observational data. The reader is referred to the original papers of the WINGS team for details about
the calculations of the theoretical luminosities, magnitudes and colors.

Occasionally, we make use of luminosities, magnitudes and colors in the same photometric system but
calculated for ideal single stellar populations (SSPs) and then extended to galaxies. The photometric data
for SSPs of different age and metallicity are taken from the Padua database of stellar tracks, isochrones,
and SSPs, magnitudes and colors in many photometric systems both in the SSP (galaxy) rest-frame as function
of the age and in the observer rest-frame as a function of the redshifts, in such a case also extensive
tabulations of the K- and E-corrections are given as  a function of the redshift for the
cosmological model of the Universe in usage
\citep{Bressan_etal_1994,Bertelli_etal_1994,Bertellietal2008,Bertellietal2009,Girardietal2002,
Girardietal2004,Tantalo_2005,Tantalo_etal_2010,Pasetto_etal_2018}. No details are given here, the reader is kindly requested
to refer to the original papers for further information.

\section{Theoretical introduction}\label{sec:2}
\subsection{Preliminary considerations on  the SRs}

{In this sub-section we provide the basic ideas generally used to interpret the observed distributions of ETGs in the SRs.
The starting point is that connected with the \Lsig\ FJ relation.
The reason is that we do not believe in a real correlation between these two variables, connecting the energetic output of stars
with their velocity dispersion, but we understand it as a consequence of the virial theorem because light traces the mass.

The virial equilibrium for ETGs can be written in this way:

\begin{equation}
    M=\frac{k_v}{G}R_e\sigma^2,
    \label{eq1}
\end{equation}
\noindent
where $M$ is the total mass of the galaxy, $k_v$ a factor taking into account the non-homology and the use
of measured structural parameters instead of theoretical quantities \citep[see for more
details][]{Donofrio2017}, $G$ the
gravitational constant, \re\ the effective radius and $\sigma$ the central velocity dispersion.
This way of writing the theorem implies that ETGs and clusters are systems
dynamically  supported by the velocity dispersion, \ie\ that all the kinetic energy is associated with the
random motion of stars/galaxies within a spherical potential (with no rotation).

If we multiply and divide by the luminosity  $L$ (in whatever band) the above expression we get:

\begin{equation}
    L=\frac{k_v}{G}\frac{L}{M}R_e\sigma^2=L_0\sigma^2
    \label{eq2}
\end{equation}

\noindent
grouping into $L_0$ the combination of mass-to-light $M/L$, \re\ and $k_v$.

As we will see in the next section the observed \Lsig\ relation has a slope of $\sim3$ and a rms scatter of
0.32, \ie\ a nearly constant proportionality factor $L_0$ valid for all systems, from the small ETGs to the big
galaxy clusters.
The virial theorem on the other hand gives $L\propto\sigma^2$ if one assume a constant $L_0$ for all systems. This means
that the combination of $M/L$, \re\ and $k_v$ should give approximately a constant value. However, if the variation in the
factor $L_0$ depends on the mass of the system, one can have a
smooth variation of $L_0$ that might cause a tilt of the \Lsig\ relation in agreement with observations, while keeping small
the scatter. This is in perfect analogy with the well known problem of the tilt of the FP \citep{Donofrio2017}.
The general impression is that the simple application of the virial theorem does not explain the FJ relation, unless one
assumes a peculiar fine-tuning among the structural parameters of galaxies.

Although eq. \ref{eq1} is formally correct, it is incomplete and imprecise because it does not
explicitly separate the mass of  the stars and gas  (baryonic mass in general) and the mass of dark matter, it does not
specify the mass-radius relationship and also neglect the possibility that other terms due to other effects are present.
To improve upon this issue, one can derive another expression for the theoretical \Lsig\
relation  based on the virial theorem developed by \citet{Caimmi2003,Caimmi2009} in which  dark
matter (DM) and baryonic matter (BM) are treated separately. The details are given in Appendix \ref{Appendix_A}.
This new \Lsig\ includes (i) a suitable relation between the star
mass ($M^*$) and the  total mass $M=M_{DM} + M_{BM}$; (ii) a suitable  relation between the stellar mass
$M^*$ and the effective radius $R_s$; and finally (iii) the redshift at which
the collapse of the proto-galaxy has taken place. In other words we take the age of the bulk stellar
population of a galaxy into account \citep[see][]{Fanetal2010}. The
relation is:

\begin{eqnarray}\label{L_sig}
\log(L) & = & 3\log(\sigma)-\log(\Gamma)-3\log(K_{\sigma})+ \\
&   & - \frac{3}{2}\log(\left( 1 + \frac{x}{y^3} \right)-\frac{3}{2}\log(1+z_f) + const,\nonumber
\end{eqnarray}

\noindent
where $\Gamma$ is the mean mass-to-light ratio of the galaxy, $K_{\sigma}$ a term that
includes the amount of DM, $x$ the ratio between the DM and BM, $y$
the ratio between the radius of the DM and BM matter components, and $z_{f}$ the redshift at the epoch of
galaxy formation.

This way of writing the FJ relation in
Eq. \ref{L_sig} indicates that, at variance with eq. \ref{eq2},  both the exponent of the \Lsig\ relation and its
proportionality factor depend on the amount of DM in the galaxies (see Appendix \ref{Appendix_A} for all details) and
are possibly variable factors.

In 2017 \cite{Donofrio2017} proposed an alternative formulation of the FJ relation that can be used to explain the origin
of the tilt of the FP and consequently the observed SRs.
Their FJ-like relation, at variance with the classical FJ relation, holds for individual galaxies and not
for a galaxy sample. In such formulation either the zero-point and the slope vary from galaxy to galaxy and depend on
the history of mass accretion and stellar evolution.
To formally distinguish it from the one of eq. \ref{eq2}, we write the
new relation in the general form:

\begin{equation}
L = L'_0 \sigma^{\beta},
\label{eq5}
\end{equation}

\noindent
where $L$ is in solar luminosities, $L'_0$ is  proportionality factor that strongly depend on the star formation history of
each galaxy and the exponent $\beta$  reflects the peculiar motion of each object in the \Lsig\ plane across the cosmic epochs.
With numerical simulations we will demonstrate in
Sec. \ref{sec:4}  that both $\beta$ and $L'_0$ are subject to variation from object to object and across
the cosmic epochs. In particular the slope $\beta$ turns out to have a spectrum of values ranging from large negative to large positive.

The new FJ-like relation hides the complex relationship existing between the baryon and DM components
and the history of mass accretion and stellar evolution experienced by each stellar system. This relation is independent of the
virial theorem. We guess that it is an equation that expresses the total luminosity of a galaxy in a way independent of the total
mass. Its existence is linked to the fact that luminosity, velocity dispersion and star formation rate are mutually correlated in
log units forming a plane as it occurs for the FP.

In the sections below using simulations we will prove that the behavior of galaxies and GCs in the \Lsig, \IeRe\ and \MR\ planes
are mutually connected and that the observed distributions that we call SRs originate from the intersection of the virial
theorem and the new FJ-like relation written for each single object.
In particular the values of $\beta$ will reproduce the main trends observed in the SRs.
 }

\section{The scaling relations of early-type galaxies and clusters} \label{sec:3}

The above introduction was aimed at clarify the framework in which we move if we want to understand the
behavior of the SRs.
In this section we start to discuss the observed distribution of galaxies and clusters in the main SRs, highlighting in particular
the comparison between ETGs and clusters.

\subsection{The \Lsig\ plane} \label{sec:2:1}

\begin{figure*}
    \centering
     {   \includegraphics[width=0.45\textwidth]{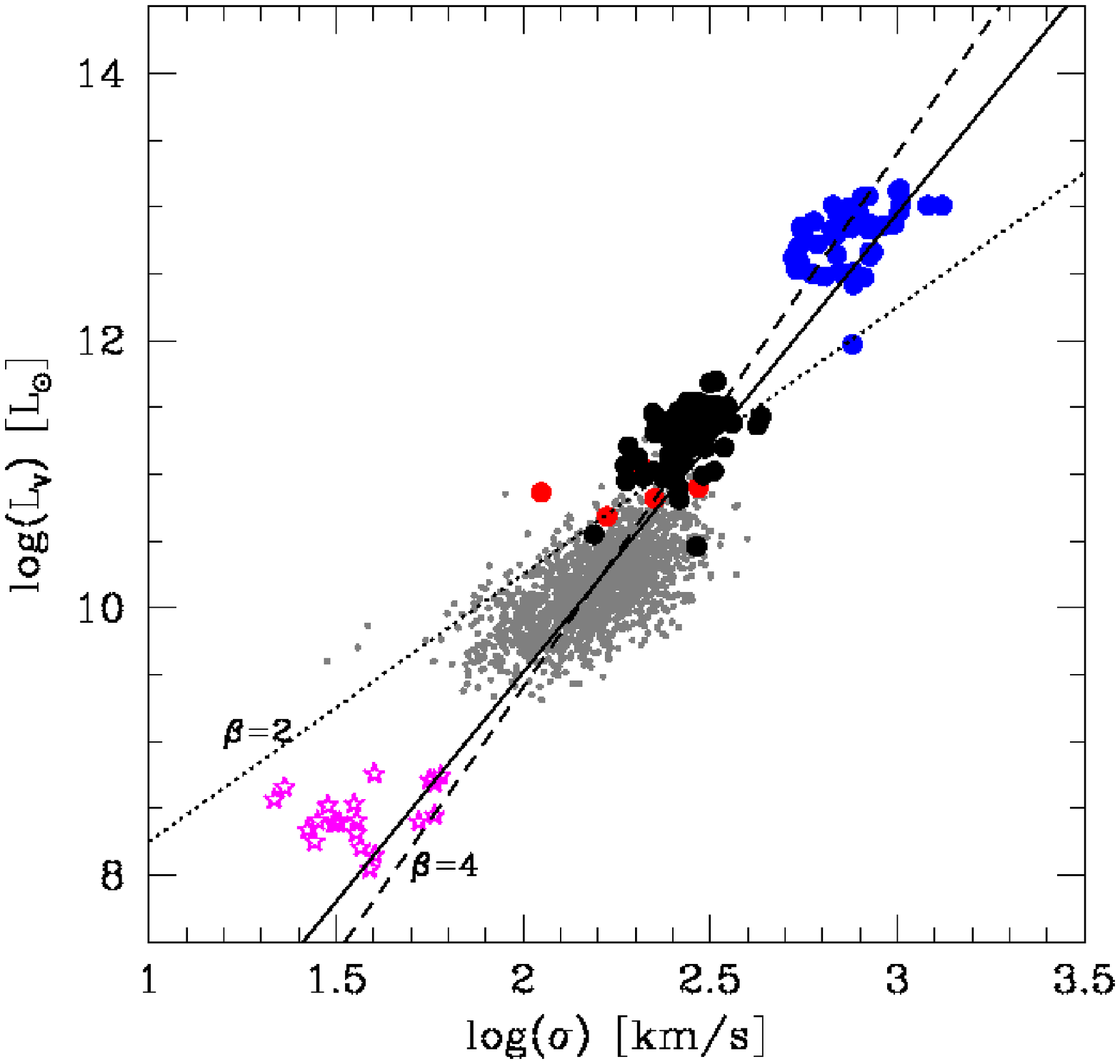}
         \includegraphics[width=0.45\textwidth]{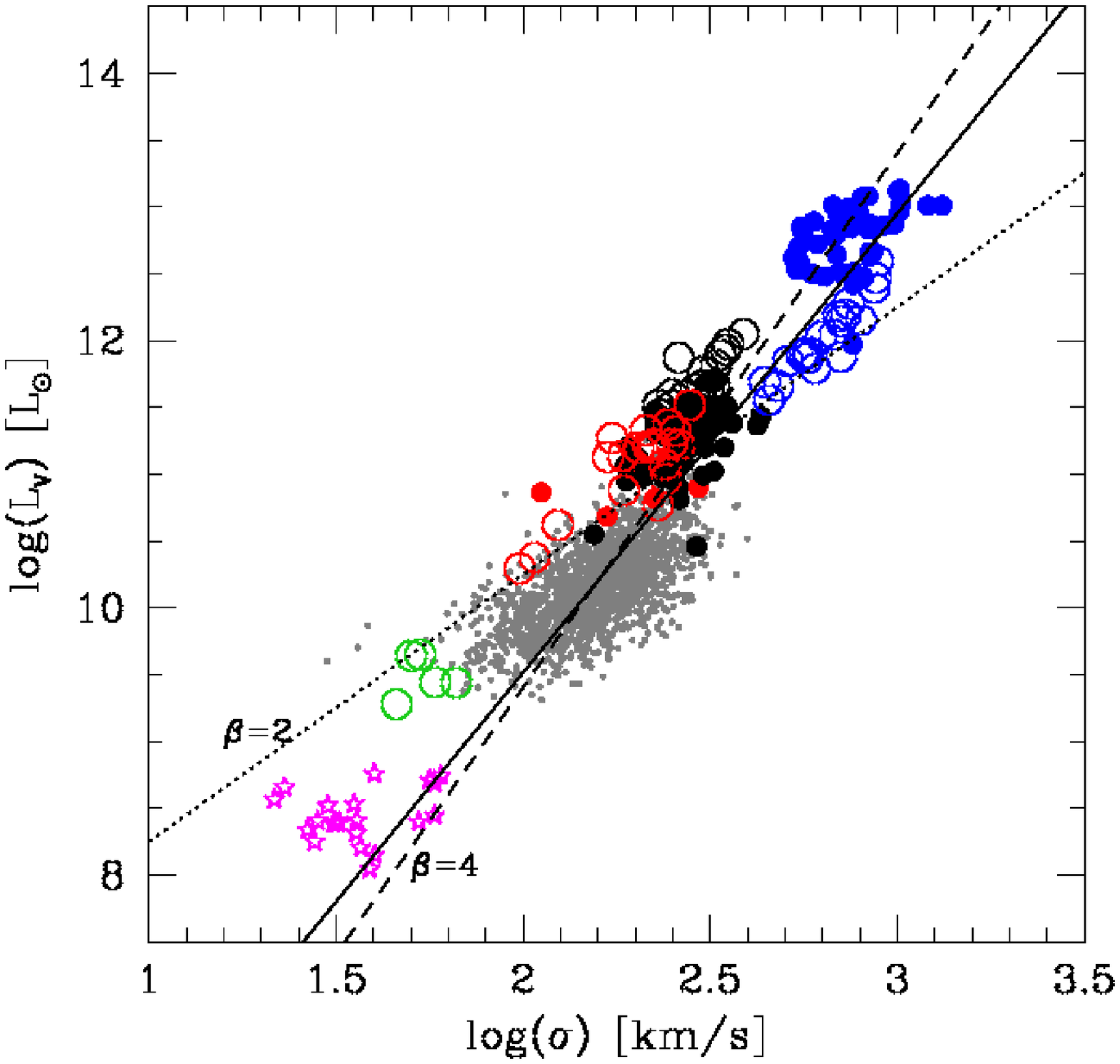} }
    \caption{Left panel: Distribution of ETGs and clusters in the \Lsig\ plane. Black filled circles mark our BCGs,
red filled circles our II-BCGs, gray small filled circles the 1729 normal ETGs used by \cite{Donofrio2017} to
study
the FP,  gray empty stars the  faint ETGs of \cite{Bettoni2016} and blue filled circles our clusters. The
normal ETGs re-analyzed in paper I are not shown because their $\sigma$ is not available. The black solid line gives the
best fit of the whole dataset obtained with the orthogonal method. The dotted line marks the $L\propto \sigma^2$ law predicted
for virial systems
while the dashed line gives the $L\propto \sigma^4$ FJ slope. Error bars are not shown because they are
approximately as big as the plotted filled circles. Right panel: The \Lsig\ plane for real and simulated
objects.
The color code and symbols are the same used before. The open circles are used for simulated objects using the
same color code. The plotted lines are the same described in
the left panel.}
    \label{fig:Lsig}
\end{figure*}

The distribution of our systems in the \Lsig\ plane is presented in the left panel of Fig. \ref{fig:Lsig}.
The data sample includes in this case the faint ETGs studied by \cite{Bettoni2016} (star symbols), the normal ETGs studied by
\cite{Donofrio2017} (small gray dots),
the II-BCGs and BCGs from paper I (red and black symbols respectively) and the GCs studied
by \cite{Cariddietal2018} (blue dots). The figure clearly shows that there is a well defined linear
trend in log scale between total luminosity and velocity dispersion for all systems, from faint
ETGs (magenta stars) to big clusters (blue dots). The solid black line in the figure marks the least square
orthogonal fit obtained by the program SLOPES \citep[][]{Feigelson1992} for the whole set of data.
{The coefficients derived with the different types of SLOPES analysis are listed in Table \ref{Tab_coef}.
The correlation coefficient is $c.c.=0.82$ and the rms scatter is 0.32.
The three boxes of the Table refer respectively to: the whole dataset of real galaxies (upper box), the subsample of normal
ETGs (middle box) and the whole sample of simulated objects (lower box).
The Table also give the errors on the parameters obtained with all the bootstrap and jackknife analyses by SLOPES.

The slope is close to the value of 4 originally proposed by \cite{FaberJackson} (shown by the dashed line).
Note that the log relation is quite linear and seems valid almost independently on the mass and size of the
systems and has approximately the same zero-point for all objects (within the observed scatter of 0.32). Note
also that the exponent is $\sim3$, \ie\ quite close to that predicted in Sect. \ref{sec:2} above,
but different from 2, the value expected for  virialized systems.

We observe that the orthogonal fits are approximately consistent with a slope $\sim 3$
for the whole set of systems and for real galaxies. A lower value is observed for simulated galaxies. In general we note that
the slopes are different when different methods are used to derive the fits.
This is due to the fact that the different samples are not of equal size. We have approximately 2000 ETGs, 60 GCs and 25 DGs.
Clearly the result of the fit is strongly affected by the distribution of ETGs.

The coefficients of the fits obtained for the single subsamples might also largely deviate from each other, in particular
for the DGs and GCs that have a clumpy distribution.

The right panel of Fig. \ref{fig:Lsig} shows the data of the Illustris simulation for the brightest
galaxies (BCGs and II-BCGs) and the faint ETGs. We can say that a qualitative good agreement with observations exists.
The agreement is poorer for GCs and faint ETGs, the former appearing systematically fainter
in luminosity and with a smaller central velocity dispersion, while the latter being a bit brighter with
respect to the observed trend of faint ETGs.
The difficulty of simulations in reproducing the properties of clusters and faint ETGs were already noted
in paper I.
Note that most of the simulated objects are approximately distributed along the slope equal 2 predicted
for virialized objects (the dotted line). This is due to the fact that the velocity dispersion of
clusters have been obtained from the virial relation, while the measured ones were calculated on the basis of
the redshift differences of the galaxies with respect to that of the central BCG.  Unfortunately the data of the simulation
does not permit an easy way to derive the GC velocity dispersions.

At this point we want to stress that our aim is not that of determining the best slope of the FJ relation, neither to quantify
the agreement between real and simulated data; we are simply comparing qualitatively the distributions of real and simulated
galaxies. We believe that the observed position of all these systems, both real and simulated, in this plane is sufficient
to agree on the fact that they all follow a quite similar trend, whatever the correct slope is.}

\begin{table}
	\begin{center}
		\caption{The coefficients of the \Lsig\ relation with the different methods provided
                by the SLOPES  program. The fitted relation is
                $log L ={\beta}'log \sigma + log L'_0$. The upper box list the coefficients
            obtained considering the whole dataset, the middle box those derived for
        the sample of ETGs and the bottom box the result obtained for simulated galaxies.}
		\label{Tab_coef}
		\begin{tabular}{|c| c| c| c| c|}
			\multicolumn{5}{c}{} \\
			\hline
			$L-\sigma$ fit   & ${\beta}'$  & $\Delta {\beta}'$ & $log L'_0$ & $\Delta
                        log L'_0$ \\
			\hline
			\multicolumn{5}{|c|}{WHOLE DATASET} \\
			\hline
			bilinear    & 2.86   &  0.06 & 3.90 & 0.14  \\
			standard  & 2.38   &  0.07 & 4.98 & 0.16  \\
			orthogonal  & 3.43   &  0.06 & 2.65 & 0.14  \\
            \hline
			\multicolumn{5}{|c|}{WINGS ETGs} \\
			\hline
			bilinear    & 2.12   &  0.05 & 5.32 & 0.10  \\
            standard  & 1.55   &  0.05 & 6.78 & 0.10  \\
            orthogonal  & 3.20   &  0.09 & 3.15 & 0.20  \\
            \hline
            \multicolumn{5}{|c|}{ILLUSTRIS} \\
            \hline
			bilinear    & 2.35   &  0.10 & 5.63 & 0.24  \\
			standard  & 2.29   &  0.10 & 5.93 & 0.27  \\
			orthogonal  & 2.47   &  0.10 & 5.40 & 0.24  \\
            \hline
		\end{tabular}
	\end{center}
\end{table}

The existence of the \Lsig\ relation has never been interpreted as a physical link between galaxy luminosity and
velocity dispersion. The common explanation for the tilted slope with respect to the virial expectation is
that there is a smooth variation of the stellar population (variation of $M/L$) and/or a smooth variation of
non-homology (variation of $k_v$ and $n$) \citep[see,][]{Donofrio2017} across the whole systems.
The reason of the mismatch is the same of the FP. In the \Lsig\ relation $L$ is used instead of the combination
of \re\ and \muem. The tilt occurs because eq. \ref{eq2} is valid for only one galaxy and not for the whole set
of ETGs. Each galaxy is in virial equilibrium, but the zero-point is different for each system.
If this interpretation is correct, the question is how the variations in structure and stellar population
occurred in the galaxies through the cosmic epochs can preserve the small scatter being $L_0$ and $\beta$
variable factors. This is the well known fine-tuning problem already encountered in the FP
\citep{Donofrio2017}.
Clearly the existence of a fine tuning between galaxy structure and stellar population is difficult to reconcile
with the idea of galaxies in continuous merging and interaction among each other that the modern numerical
simulations have shown.

In Sec. \ref{sec:4} we will demonstrate through simulations that the current slope of
this relation originate from the global complex mass assembly history of galaxies, that clearly affect either
the mass-to-light ratio and the structure of the systems.

\begin{figure*}
    \centering
     {   \includegraphics[width=0.45\textwidth]{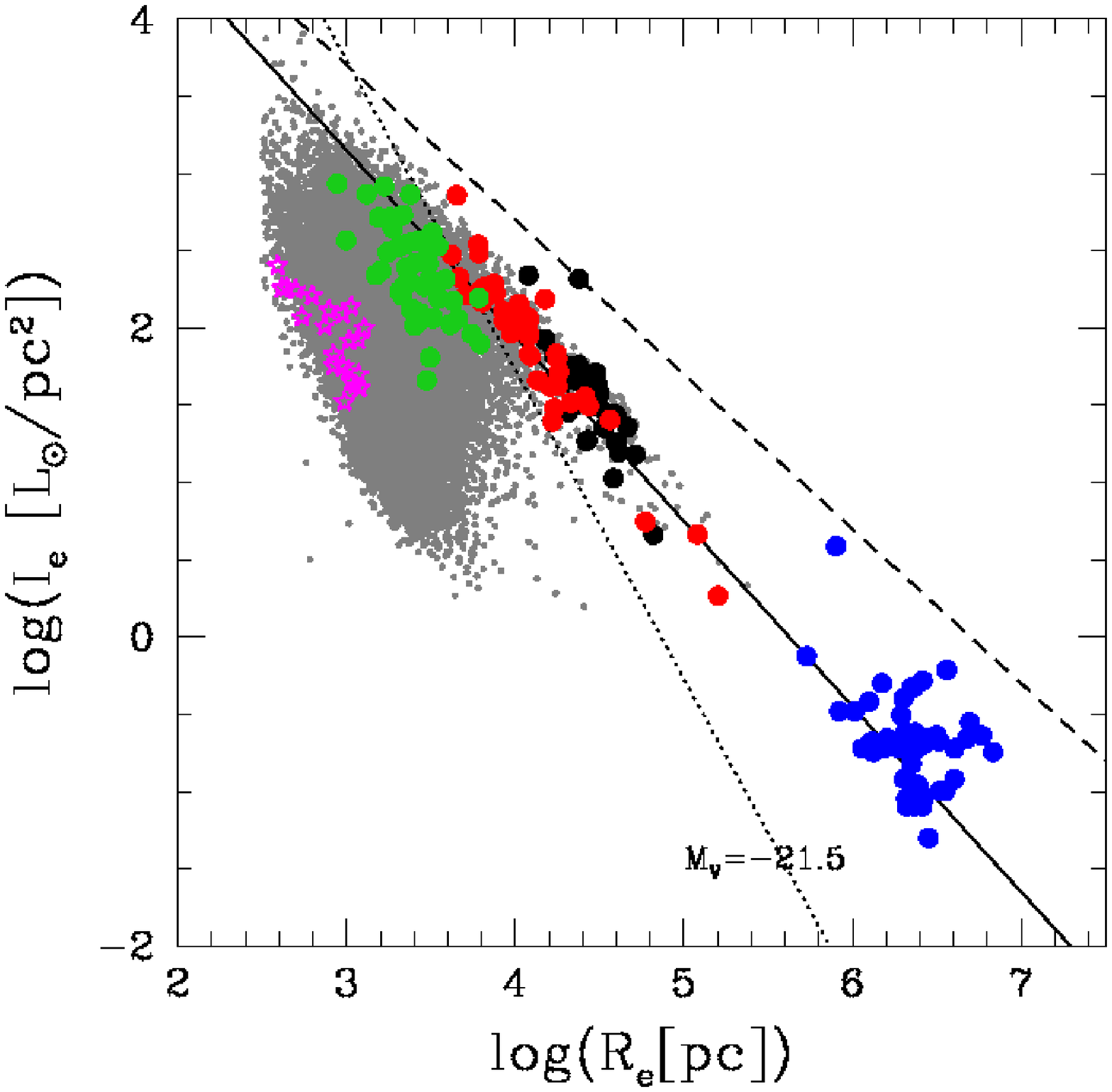}
         \includegraphics[width=0.45\textwidth]{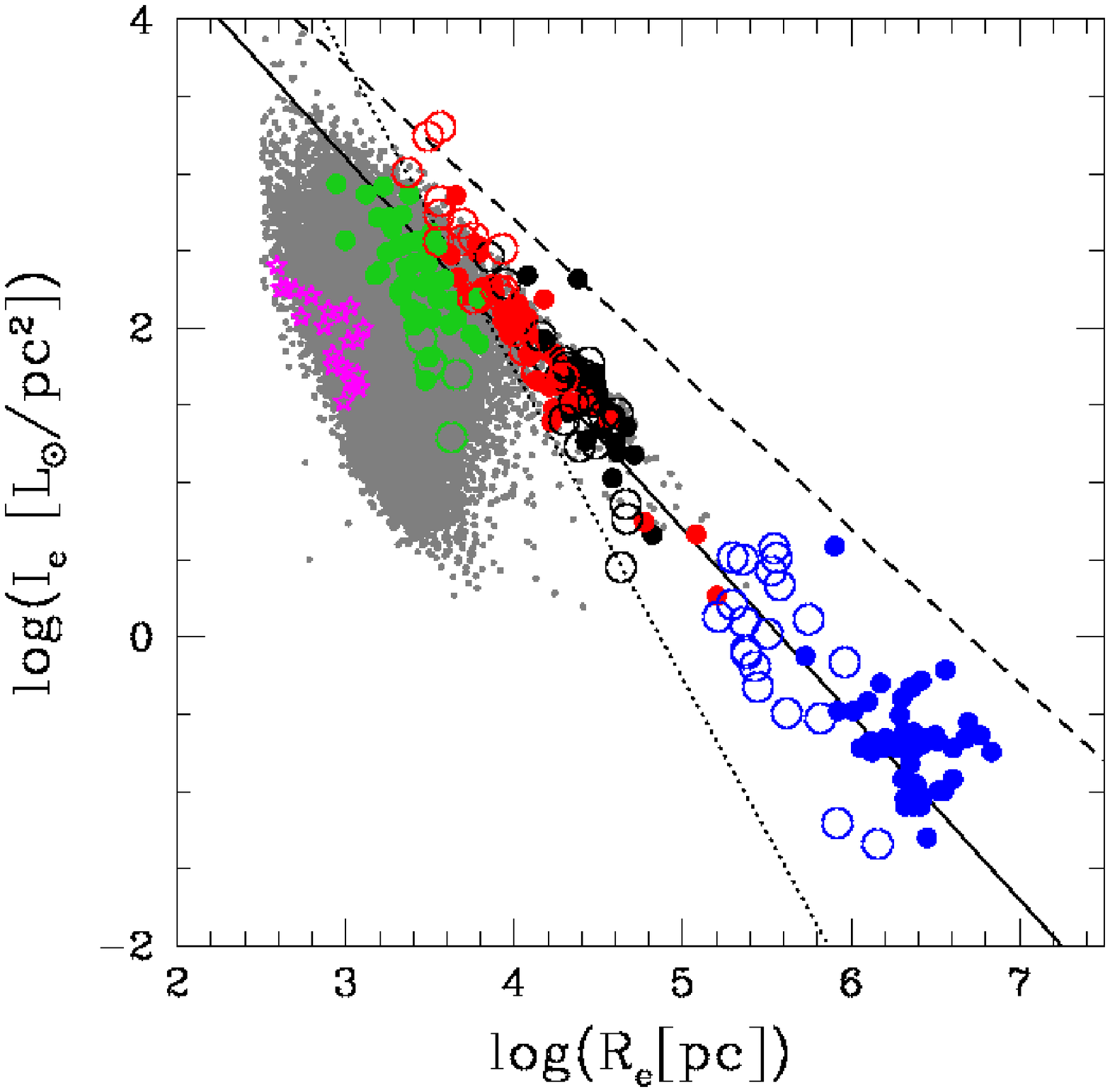} }
    \caption{Left panel: Distribution of galaxies and clusters in the \IeRe\ plane. Black filled circles mark
our BCGs,
red filled circles our II-BCGs, green filled circles our random sample of normal ETGs, gray filled dots are
the 34982 galaxies analyzed by \cite{Donofrio2014} with GASPHOT,
empty magenta stars are the faint ETGs of \cite{Bettoni2016} and filled blue circles our clusters. The dashed
line with slope $-1$ is that predicted for virialized systems for a possible ZoE. The zero-point of this line
has been chosen as explained in Sec. \ref{Zoe}. The dotted line is that expected for systems of equal
luminosity $M_V=-21.5$ with slope $-2$. The solid line with slope $-1.2$ is that obtained by
\cite{Capacciolietal92}. Right panel: The \IeRe\ plane with real and simulated objects. The color code and
symbols are the same used before. The open circles with the same colors of real galaxies are used for
simulated objects. The plotted lines are the same of the left panel.}
    \label{fig:IeRe}
\end{figure*}

\begin{figure*}
    \centering
     {   \includegraphics[width=0.45\textwidth]{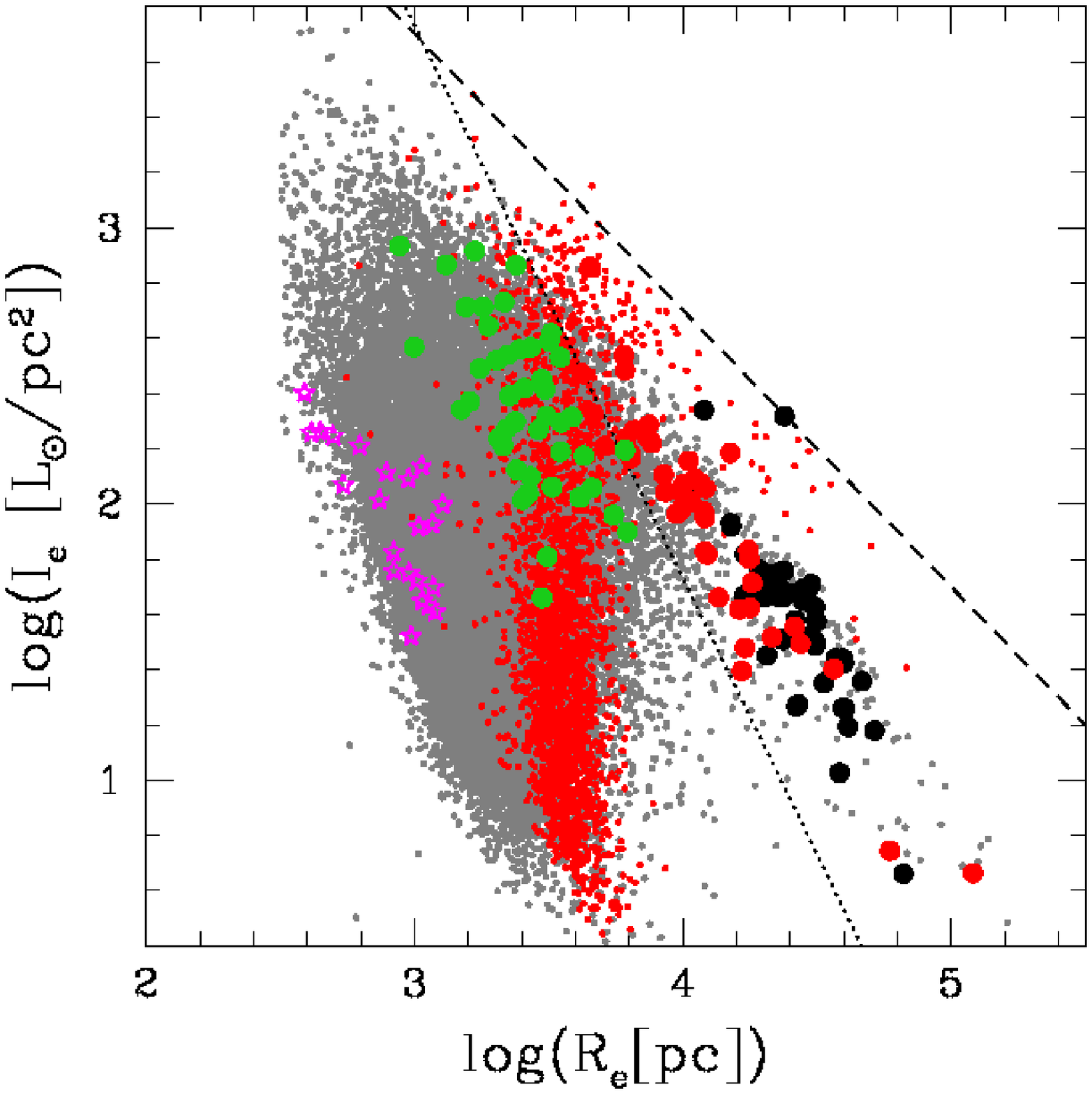}
         \includegraphics[width=0.45\textwidth]{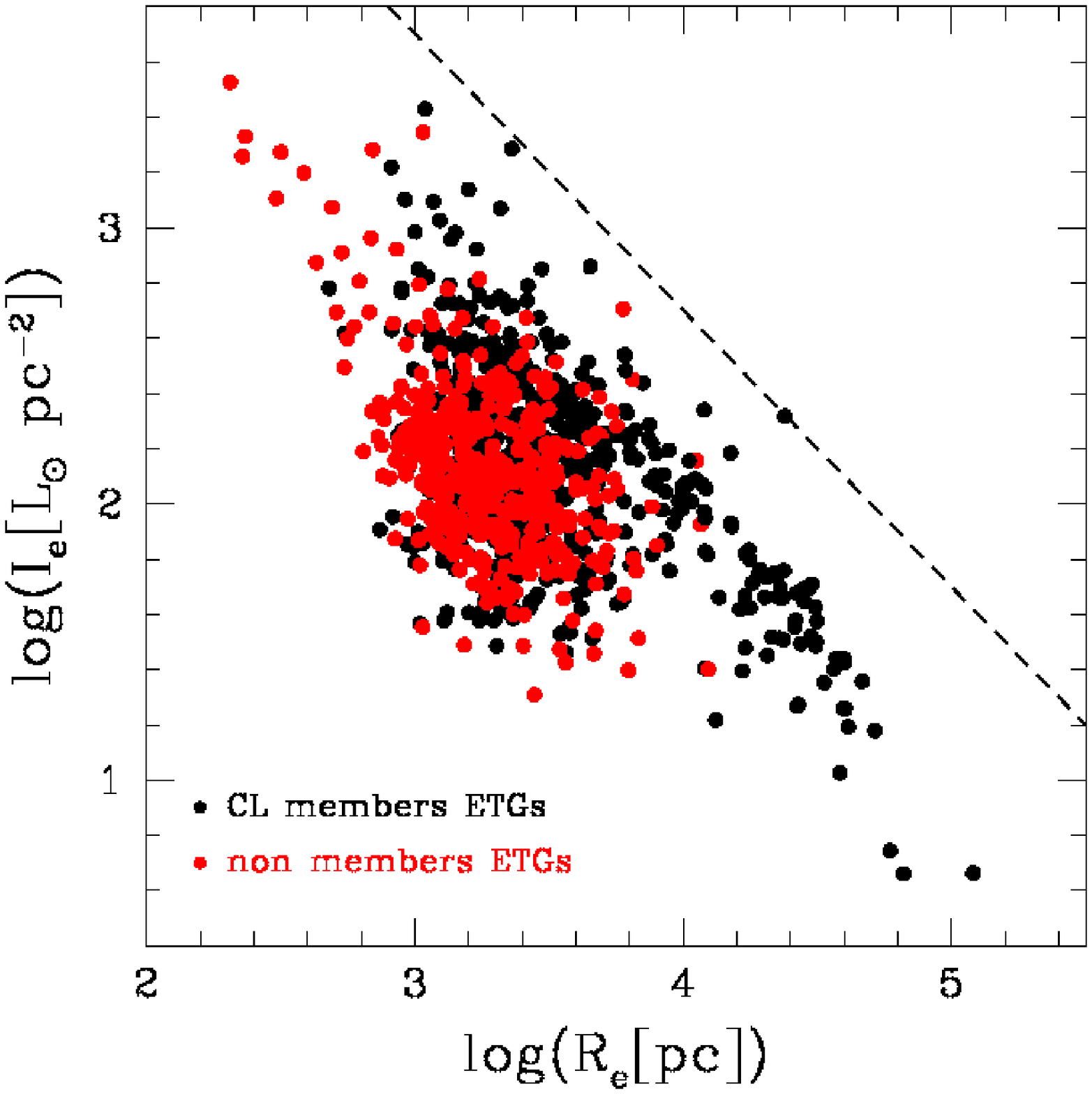} }
    \caption{Left panel: enlargement of the \IeRe\ plane with real and simulated galaxies. Black filled
circles
mark our BCGs,
red filled circles our II-BCGs, green filled symbols our random sample of normal ETGs, gray filled dots are
the WINGS ETGs of \cite{Donofrio2017},
empty stars are the faint ETGs of \cite{Bettoni2016} and filled circles our clusters. The small red dots are
 used for the whole set of Illustris galaxies at $z=0$. In this case the effective mass radius has been
assumed to be equal to the effective radius. The dotted line is that expected for systems of luminosity
$M_V=-21.5$. Right panel: The \IeRe\ plane for cluster (black dots) and non-cluster (red dots) ETGs. Here we
have used the sub-sample of galaxies with available masses from \cite{Fritzetal2007}. In both panels the
dashed lines are the trends for virialized systems with slope $-1$ and a zero-point of a possible ZoE.}
    \label{fig:IeReB}
\end{figure*}

\subsection{The \IeRe\ plane} \label{sec:3:2}

\cite{Kormendy1977} first recognized that the distribution of ETGs in the \IeRe\ plane is not random and that
the slope of the observed distribution is not that predicted for simple virialized systems.

Remembering eq. \ref{eq2} and using the definition of surface brightness we can write:

\begin{equation}
    \langle I \rangle_e=\frac{k_v}{2\pi G}\frac{L}{M}\sigma^2R_e^{-1}
    \label{eq6}
\end{equation}
\noindent
so that in log units the slope of the virial \IeRe\ relation is $-1$. For systems along this line (i.e. with
the same zero-point and similar $k_v$) the mass-to-light ratio $M/L$ should scale with $\sigma$ according to $M/L\propto\sigma^2$.

\begin{figure}
    \centering
	\includegraphics[width=0.45\textwidth]{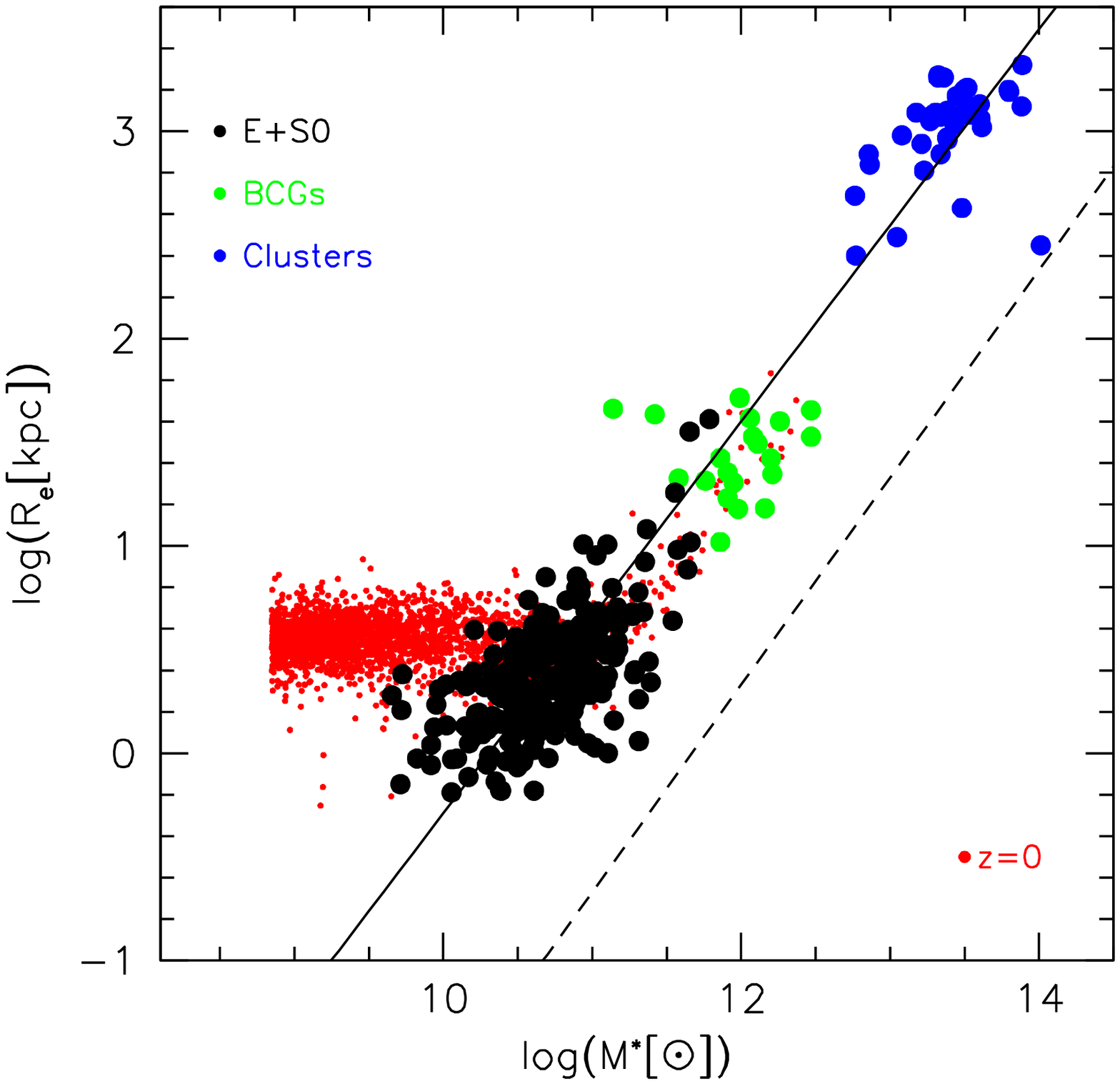}
    \caption{The distributions in the \MR\ plane for normal ETGs (black filled circles), BCGs
(green filled circles) and clusters of galaxies (blue filled circles) from our WINGS samples. The red small
dots mark the data of the Illustris simulations for galaxies at $z=0$. The solid line is the fit of the
galaxies and cluster sample, while the dashed line is the slope predicted from the virial theorem for a
possible ZoE.}
    \label{fig:MR_BCGCL}
\end{figure}

The left panel of Fig. \ref{fig:IeRe} shows how our galaxies and clusters are distributed in such plane.
Observe that the distribution of BCGs, II-BCGs, normal ETGs, faint ETGs and clusters do not follow
the slope predicted by the virial theorem, but a much steep trend (look at the solid line between the dotted
line predicted for systems of
equal luminosity with slope $-2$ and the dashed line). This is the line found by \cite{Capacciolietal92} best
fitting a much larger sample of
bright ETGs. The slope is $-1.2$ (that in surface brightness units is 3; \muem$=3.0\log$(\re[kpc])). In
their work \cite{Capacciolietal92}
distinguished two different families of ETGs in this plane: the `ordinary' family with faint luminosity and
small radii, and the `bright' family with high luminosity and large radii. These two families are distributed in a
completely different way in the \IeRe\ plane, probably for the different role of merging in their formation.
The 'ordinary' family is well visible with the present sample in Fig. \ref{fig:IeRe}: it is made
by objects with \re$\leq4$ kpc (the green dots, the open magenta stars and the green filled circles).
\cite{Donofrio2014} already showed that spiral galaxies are confined to the 'ordinary'  family (their Fig. 9).
The figure clearly indicates that only the brightest ETGs develop the tail well known as the Kormendy's
relation.

The ZoE is the region empty of points above the dashed line with slope $-1$ for virial systems.
We will see in Sec. \ref{Zoe} how the zero-point of this line has been obtained.

\begin{figure}
    \centering
     {   \includegraphics[width=0.45\textwidth]{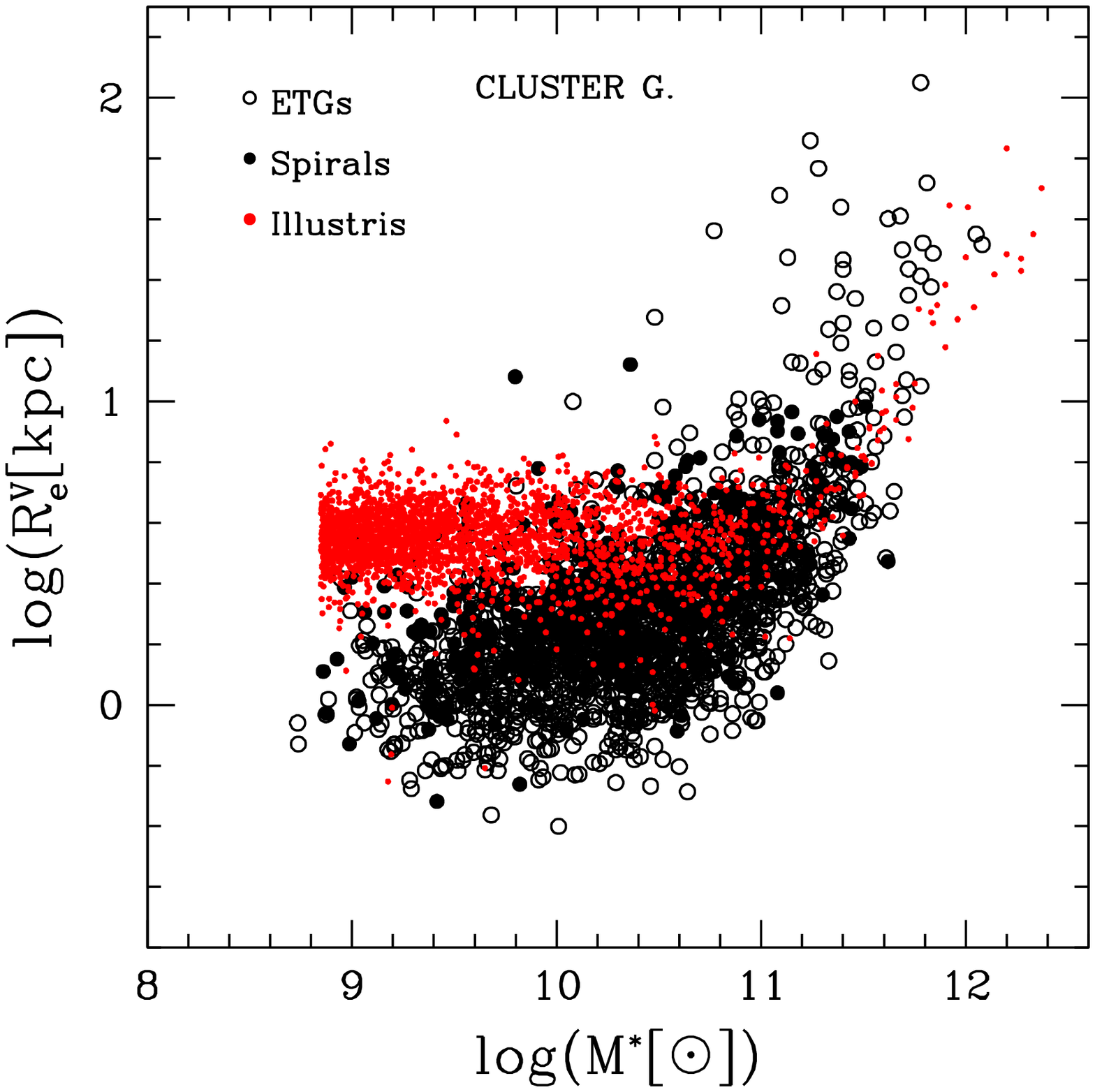}
         \includegraphics[width=0.45\textwidth]{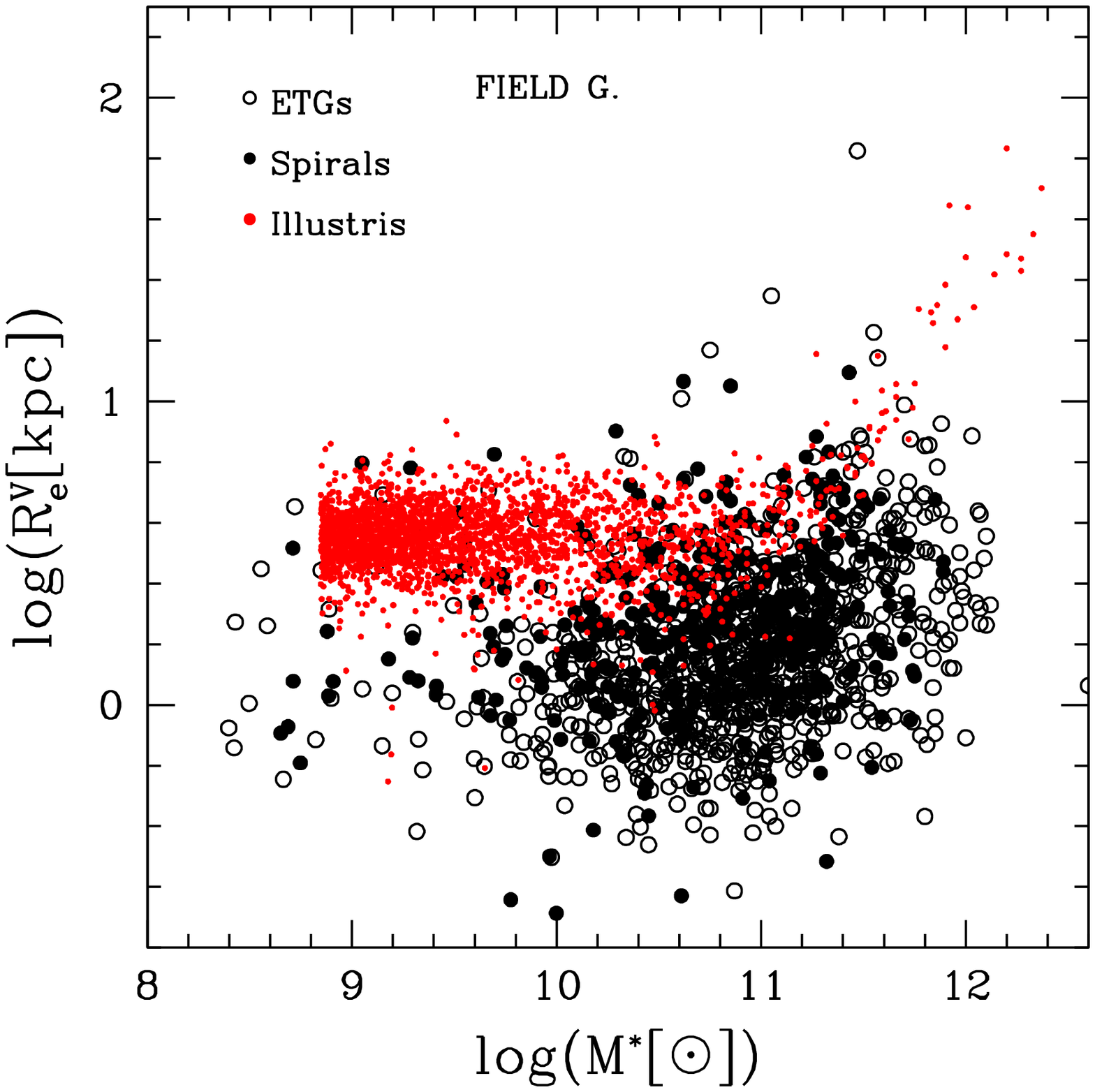} }
    \caption{The stellar Mass-Radius relation for galaxies in clusters (upper panel) and in the field
(bottom panel). The open black circles mark the real ETGs, the black filled circles the spiral galaxies, and
the red dots the simulated data at $z=0$. The stellar masses used here have been derived from the K-band
luminosity of our galaxies.}
    \label{fig:MR}
\end{figure}

In Fig.\ref{fig:IeRe} we see that clusters share the same properties of big ETGs. Their position is at
low surface brightness and large radii along the line fitting the high luminous galaxies. Clusters therefore
follow the same \IeRe\ relation of bright ETGs.

In paper I, when we compared the light profiles of clusters and ETGs, we concluded that clusters are more
similar to faint ETGs than to BCGs. Here instead we see the opposite. When we consider the structural
parameters they are more similar to BCGs. We will attempt a possible explanation of this behaviour in Sec.
\ref{sec:5}.

\begin{figure*}
    \centering
     {   \includegraphics[width=0.45\textwidth]{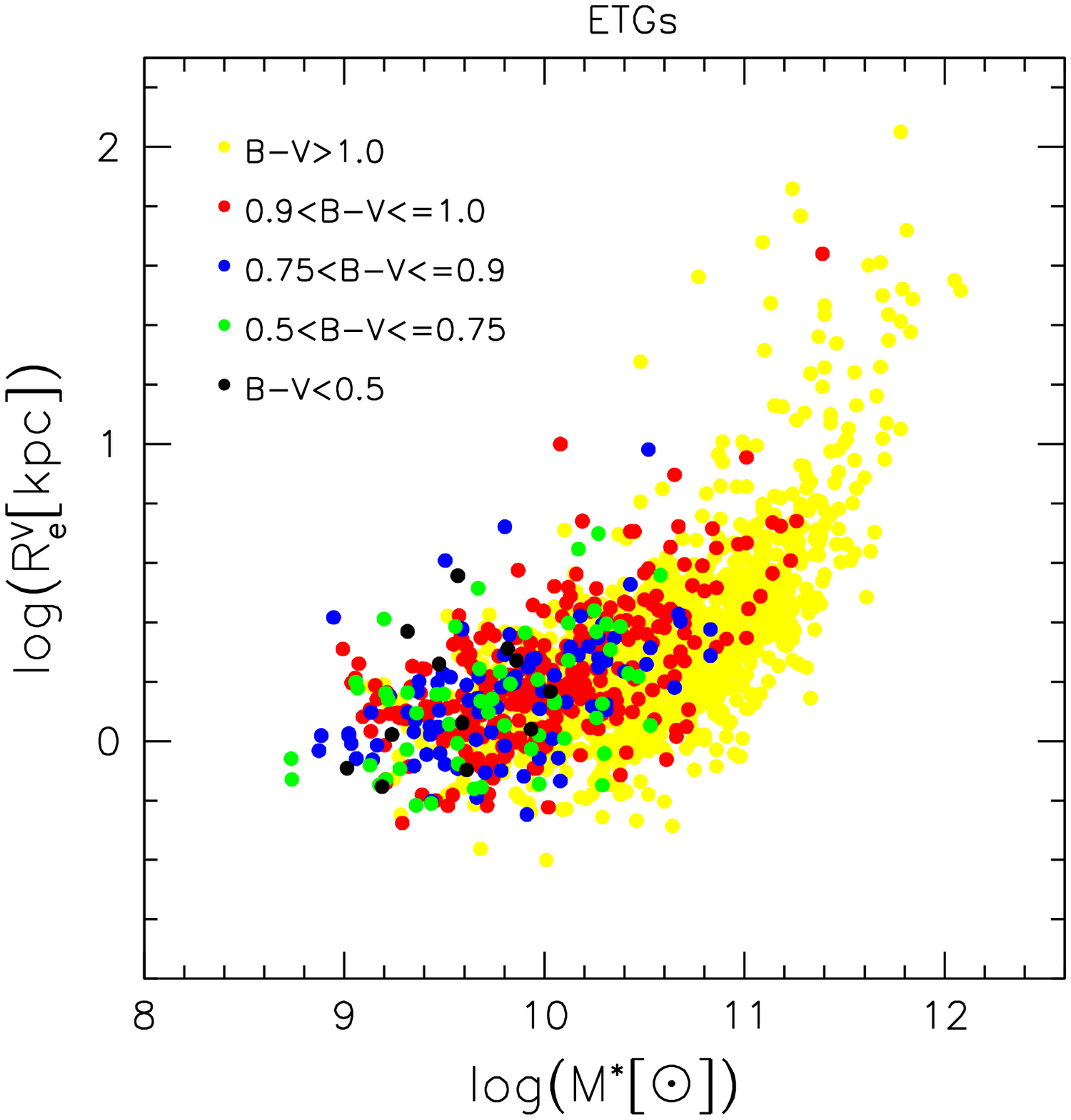}
         \includegraphics[width=0.45\textwidth]{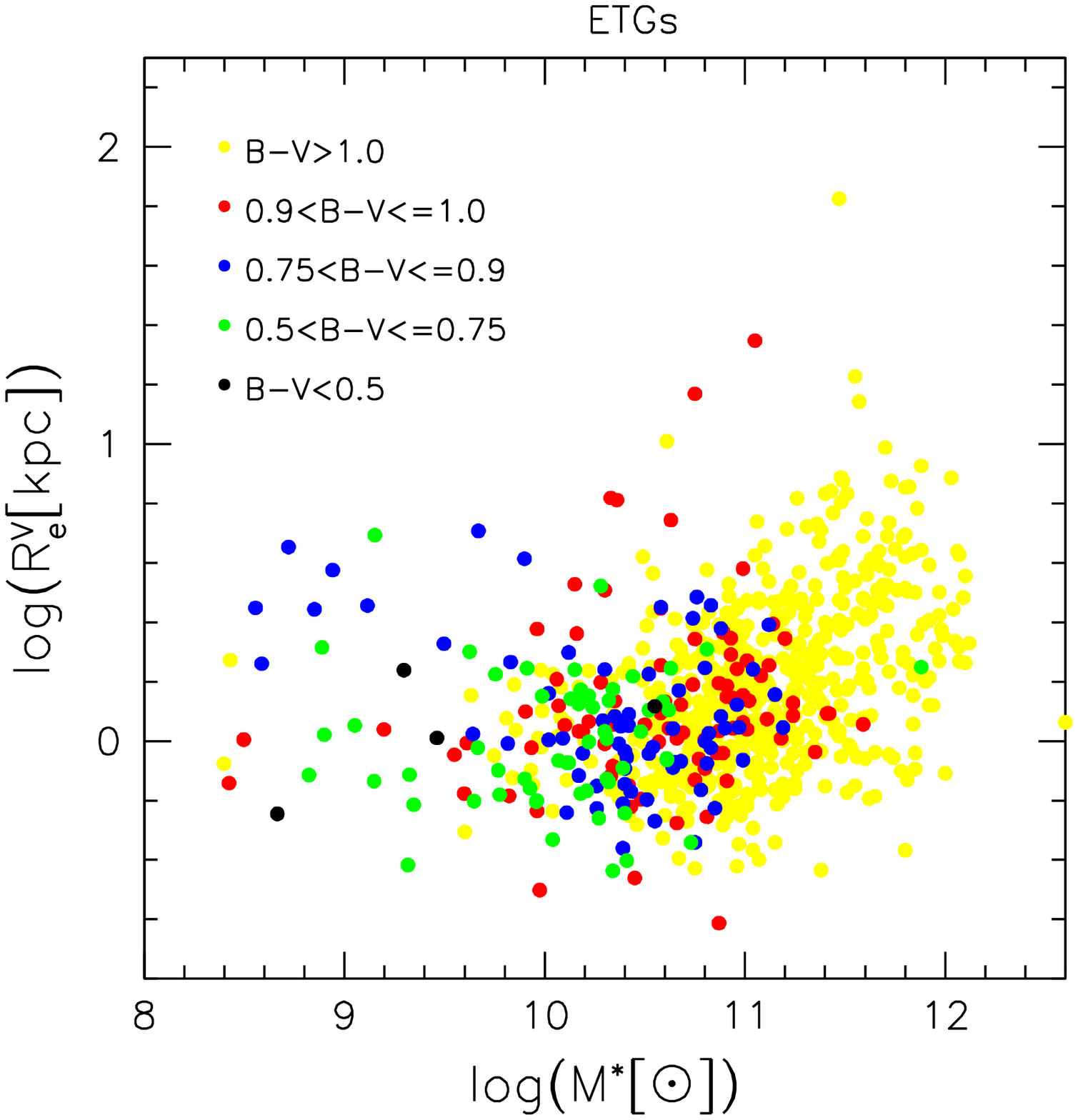}
         \includegraphics[width=0.45\textwidth]{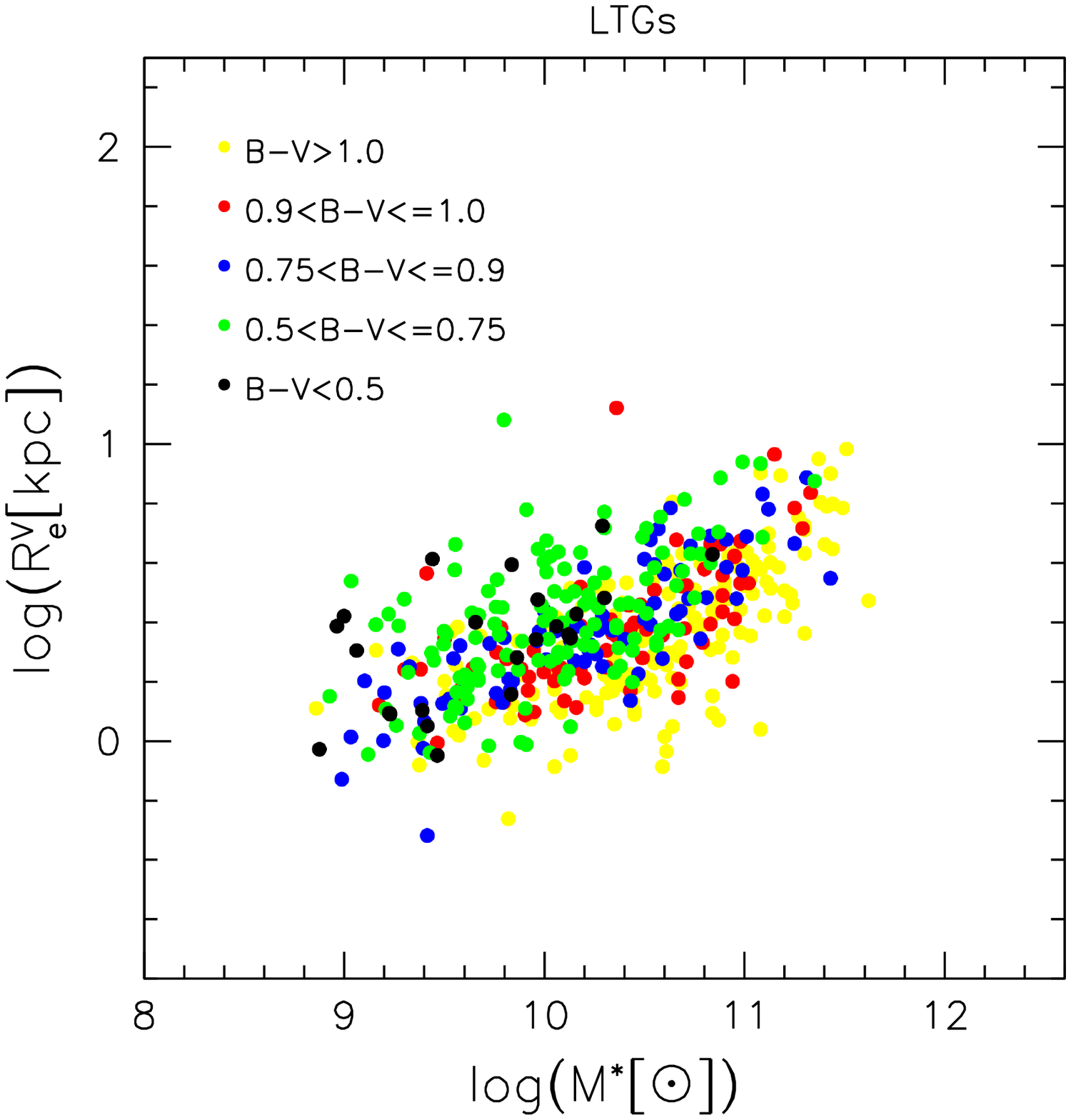}
         \includegraphics[width=0.45\textwidth]{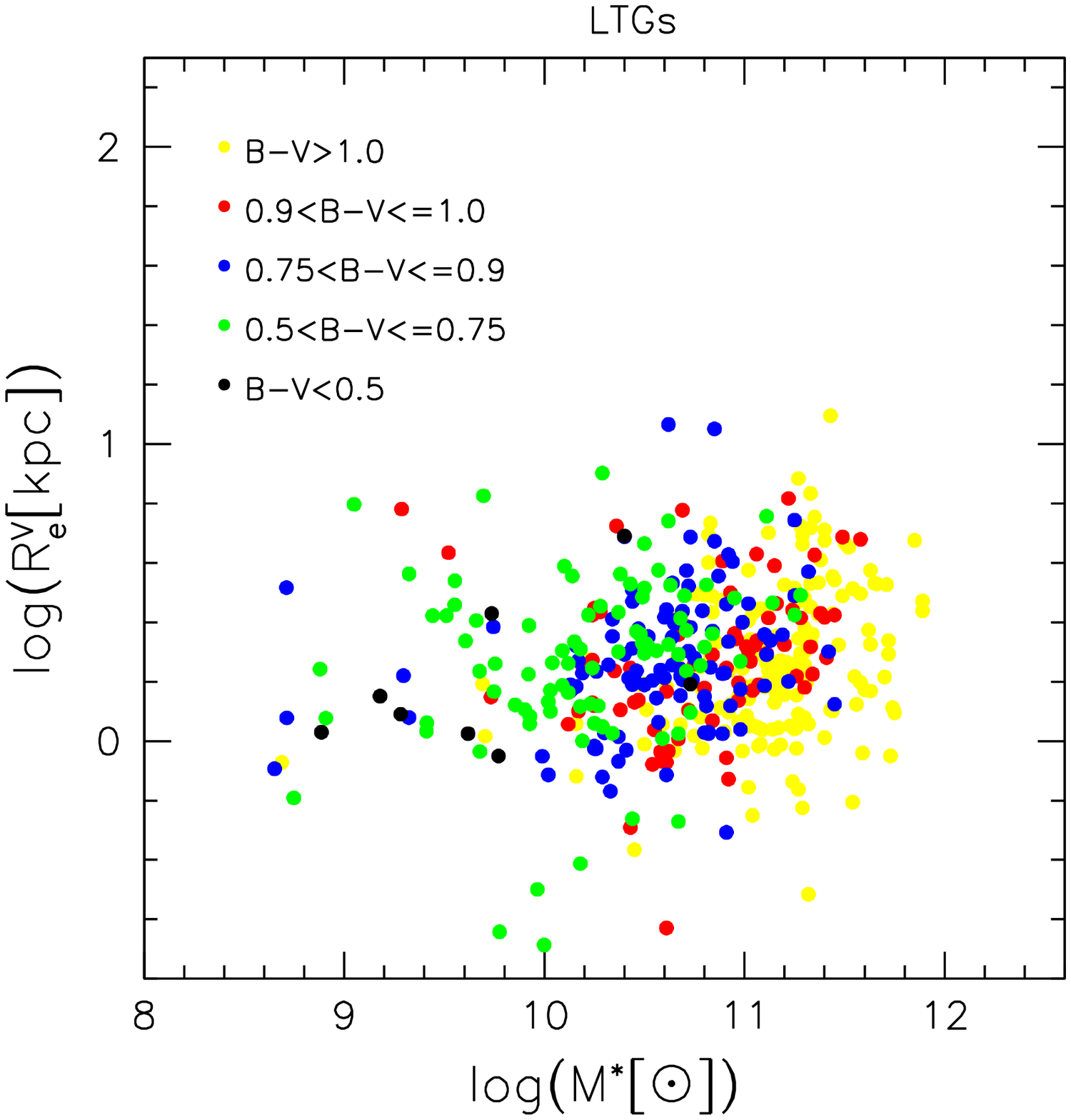}  }
    \caption{The \MR\ relation for ETGs and LTGs in clusters (left panels) and in the field (right panels).
Galaxies are plotted with different colors according to their $B-V$ color index.}
    \label{fig:MRBmV}
\end{figure*}

Now look at the right panel of Fig. \ref{fig:IeRe} showing the \IeRe\ plane with simulated data.
For each simulated object at $z=0$ we have derived the growth curve luminosity profile and the main structural
parameters (\re, \Ie, $\sigma$, etc.) following the same procedure used for real galaxies. We can therefore compare
the position of the simulated structural parameters (open circles) with the real ones.
The good agreement achieved by simulations for BCGs, II-BCGs, and normal ETGs, and the failure for clusters is evident.
Clusters are systematically smaller in size and brighter in surface brightness. This confirms what we claimed in paper I.

The left panel of Fig. \ref{fig:IeReB} shows an enlargement of the \IeRe\ plane in the area covered by galaxies.
Note how the simulated data for the whole set at $z=0$ marked by the small red dots are able to reproduce
both the 'ordinary' and 'bright' family defined by \cite{Capacciolietal92} (their Fig. 4). The simulations fail
only in the zero-point of the surface brightness that appears systematically brighter than that of real galaxies. This
effect is not visible in the right panel of Fig. \ref{fig:IeRe}, because in that case the effective radius and the
effective surface brightness were obtained from our careful analysis of the light profiles of BCGs and II-BGCs done
in paper I, while here we have used the half-mass radius of the Illustris dataset
that might be a bit different from the effective radius.
The simulations seem also to fail in the effective radius of the faint ETGs, that appear systematically
bigger with respect to that of real
objects (marked by the small gray dots, the empty magenta stars and the green filled circles).

The right panel of Fig. \ref{fig:IeReB} shows the \IeRe\ plane for cluster and non cluster ETGs. In this case
we have used a sub-sample of the WINGS galaxies, the one with available masses derived by
\cite{Fritzetal2007}. Note how the tail of galaxies with large \re\ is present only for cluster objects
(marked by black dots), while is almost absent for field objects (red dots).

\cite{Capacciolietal92} attributed the origin of the 'bright' family to mergers. The data therefore seem to
suggest that in the cluster environment galaxies experience more merging events.
The big number of minor dry merging events and the stripping phenomena could in fact inflate the radius of
ETGs in particular in the central region of the clusters \citep[see e.g.][]{Naab2009}.

{\cite{Donofrio2017} showed that using eq. \ref{eq5} with negative values of $\beta$ it is possible to fit
the observed distribution of the bright ETGs in the \IeRe\ plane, \ie\ to obtain the Kormendy relation
\citep{Kormendy1977}.
This occurs because one can define two intersecting planes for each object in the 3D \muerespace\ space:
one representing the mass of the galaxy (through the virial equation) and one representing the luminosity
(provided by the \Lsig\ relation). The intersection between these planes generates a line in the \muerespace\
space that can be observed projected in the \IeRe\ plane. When $\beta$ is negative it is possible to fit the
distribution of the 'bright' ETGs and clusters. The slope of this line in the \IeRe\ plane is given by eq. 17 in
\cite{Donofrio2017}, that we rewrite here:

\begin{equation}
\log(\langle I_e \rangle) = \frac{(2/\beta)-(1/2)}{(1/2)-(1/\beta)} + \Pi
\end{equation}

\noindent
where $\Pi$ is a factor that depends on $k_v$, $M/L$ $\beta$ and $L_0$.
Table \ref{beta_values} gives for each possible value of $\beta$ the corresponding slopes in the \IeRe\
(and \muere) relation. These slopes represent the direction of motion of a galaxy in this space
along a cosmic time interval.
Note how progressively large negative values of $\beta$, that are peculiar of galaxies in a quenched state, determine values
of the slope in the \IeRe\ plane converging toward the expected virial value of $-1$.
The luminosities of these galaxies is progressively decreasing at nearly constant velocity dispersion,
a behavior of objects in passive stellar evolution.

This means that the ZoE is not only the locus of undisturbed virialized galaxies, but also that of purely
passive evolving systems. Notably this slope does not depend on the mass of the system and is the same for all types of
objects, from stars to galaxy clusters. The zero-point of the ZoE on the other hand depends on the
mass-to-light ratio and the non-homology reached by systems when they arrive to the condition of passive
evolution and virialization. As galaxies get older their $M/L$ ratio tend to increase asymptotically providing
a maximum possible value for all stellar systems.
In the \vfilt\ band the maximum measured stellar mass-to-light ratio is $\sim20$.
Young objects cannot cross this boundary limit. This might suggest that the undisturbed
virialization of galaxies can be reached only when systems enter in the passive evolution. In this condition, when no
more energy is injected in the galaxy from star formation, AGN and SN feedbacks, the system can relax and enter progressively
in the trend predicted by the virial theorem. Clearly its final position in those planes will depend on the zero-point
reached when these conditions are met.}

\subsection{The \MR\ plane} \label{sec:3:3}

When a galaxy is in the virial equilibrium one might expect that the stellar mass scales linearly (with slope 1 in log units)
with the effective radius as in eq. \ref{eq1}.

In Fig. \ref{fig:MR_BCGCL} we observe the distribution of normal galaxies, BCGs and GCs in the \MR\ plane. We see that
clusters (blue dots) follow the same distribution of BCGs (green dots) and
ETGs of mass greater than $10^{10} M_\odot$ (black dots). Remember that we have used only the clusters that
are well fitted by the \Sers\ law, \ie\ those much closer to a virial equilibrium not disturbed by secondary
components likely due to recent merging events. The red small dots are the data coming from Illustris.
The solid line best fitting this distribution of galaxies and clusters has a slope of $\sim0.9$, very close
to the value of 1 coming from the virial theorem (shown by the dashed line) that here
represents also the ZoE of the \MR\ plane. On the right of this line there are no objects.
The zero-point of this relation is discussed in Sec. \ref{Zoe}. {The same figure shows with red dots the distribution
obtained for the simulated galaxies. Note that the simulation catches the high mass tail, while it fails for the low masses.}

Fig. \ref{fig:MR} shows the stellar mass-radius relation derived only for the WINGS galaxies. Here we used
the whole set of WINGS galaxies with available stellar masses mentioned in the Introduction that was
calculated using the \kfilt\ luminosity as a proxy. We have only distinguished the various galaxies on the
basis of the membership and the morphology (ETGs and LTGs).
Cluster member objects are plotted in the upper panel and non-member galaxies in the bottom panel.
 ETGs are marked by open circles, while LTGs by filled circles. The membership was evaluated by
\cite{Cava2009} on the basis of the redshift and the morphology by \cite{Fasanoetal2012}. Note that the tail
in the \MR\ plane is primarily due to massive ETGs and is almost absent for
spirals and for field objects. Is this behaviour due to a selection bias?
The distribution in redshift of the field sample peaks at $\sim0.1$, while that of clusters at $\sim0.05$.
This is a potential source of bias for the present comparison, but simulations have revealed that \re\ does
not change significantly in this redshift interval.
The interval of mass is also quite similar, so that we can be quite confident that the observed difference is
not originated by selection effects. In addition we know that the galaxies with the largest radii are also the
more luminous, so that we can exclude a Malmquist bias.

In the two panels the data of the Illustris dataset at $z=0$, derived only for the galaxies in clusters, are again shown
by small red dots.  The banana-like shape  of the distribution of real
galaxies is not well reproduced. In the Illustris-TNG the effective radii are a bit lower, but it is seems
that they are on average still too high by a factor of 3 \citep[see Fig. 1 of][]{Genel2018}.

We have verified that the galaxies in the tail are the same observed in the tail of bright galaxies in
Fig. \ref{fig:IeRe}. The tail is formed primarily by massive quenched objects at the center of clusters,
that have likely increased their radius for the frequent dry merger events. On the right of this tail there
are no galaxies. This is the ZoE region of the \MR\ plane.
We will show below that the slope followed by massive quenched passive objects in this diagram is the same
of that predicted for virialized systems.

Fig. \ref{fig:MRBmV} shows the distribution of ETGs and LTGs in clusters and in the field using different
colors for the different ranges of the $B-V$ index of galaxies. Note that red objects are preferentially
distributed in the right part of the diagram, \ie\ are closer to the ZoE. Furthermore, the banana shape is
more evident for ETGs than for LTGs. The trend is almost absent for LTGs in the field, while for objects in
clusters the relation is always present. The LTGs in clusters seem to share a \MR\ relation not present in
the field. Again we are led to think that even LTGs grow in size in the cluster environment. A very similar
trend is seen when different ranges of the S\'ersic index $n$ are considered. This means that
the structure of the galaxies also changes along the sequence: high values of the S\'ersic index are measured
only for the galaxies in the tail, while low values of $n$ are typical for the flat part of the sequence.

Table \ref{Tab:MR_fit} shows the coefficients (slope and intercept) of the best fit linear relation for the
galaxies distribution in the \MR\ plane, when different ranges of masses are selected. The best fit relation has
been obtained with the standard least square fitting technique
\citep[using the program SLOPES of][]{Feigelson1992}.
It is clearly visible that the slope increases when massive galaxies are taken into account: we start from 0.13
(when low mass systems are fitted) and we end up with 0.68 (when only the most massive systems are fitted).
The slopes of the fit changes a little bit if the bilinear least square fit is applied, reaching values up
to 1, when the fit is done only for the massive galaxies ($\log(M^*>10.5)$).
The average errors on the slopes and intercepts are of the order of 0.02 and 0.2 respectively. This means that
the observed differences are significant.

\begin{table}
\begin{center}
\caption{Slope and intercept of the best fitted \MR\ relation for different mass ranges.}
		\label{Tab:MR_fit}
		\begin{tabular}{|c| c| c|}
\multicolumn{3}{c}{} \\
\hline
         $M^*$ range    &  slope  & intercept \\
\hline
         $M^* \leq10^{10.5}$      &     0.13     &      -1.10   \\
         $M^* \leq10^{11.5}$      &     0.28     &      -2.42   \\
         $M^* \leq10^{12.5}$      &     0.30     &      -2.77   \\
\hline
\multicolumn{3}{c}{} \\
\hline
         $M^*$ range    &  slope  & intercept  \\
\hline
         $M^* \geq10^{9.5}$       &     0.34     &      -3.27   \\
         $M^* \geq10^{10.0}$      &     0.45     &      -4.44   \\
         $M^* \geq10^{10.5}$      &     0.68     &      -6.91   \\
\hline
\end{tabular}
\end{center}
\end{table}

This behavior demonstrates that the distribution of galaxies in the diagram is curved. Probably the origin
of the trend should be searched in the different conditions of virialization and density distribution inside
the single galaxies. The pure virial behavior of eq. \ref{eq1} with a similar zero-point seems to be valid
only for the most massive and red systems. In less massive ETGs and in LTGs rotation is progressively more
important, as well as the DM content.
It is also possible that dwarf systems are not in a full virial equilibrium yet, being still affected by
episodes of star formation and in general suffering the interactions with the cluster environment (stripping
and harassment).
They might be not fully relaxed from an energetic point of view, presenting a radius much larger than that
expected for a virial system of that mass.
These two effects could be at the origin of the curved distribution of the \MR\ relation. We have better
analyzed this relation in paper III of this series \citep{Chiosietal2019}.
We only want to note that the observed distribution reveal a systematic change of zero-point of the
virialized galaxies. Such variation has been also invoked  for explaining the tilt of the FP and FJ relation,
and the observed distribution in the \IeRe\ plane tilted with respect to the virial prediction.

{ Now we want to show  what happens when we combine eq. \ref{eq1} with eq. \ref{eq5}. A simple algebra gives:}

\begin{equation}
R_e = \left(\frac{1}{\frac{k_v}{G}\left(\frac{2\pi\langle I_e \rangle}{L'_0}\right)^{2/\beta}}\right)^{1/\left(4/\beta+1\right)}
M^{1/\left(4/\beta+1\right)}.
\label{eqMR}
\end{equation}

{In Table \ref{beta_values} we have listed the values of the predicted slopes for the \MR\ relation on the basis of the
possible values of $\beta$, \ie\  once the virial plane and the FJ-like relation are combined. Note how the slope of the
\MR\ relation is in agreement with the values fitted on the observed distribution provided in Table \ref{Tab:MR_fit} below.
The resulting curved distribution is clearly obtained by the progressive change of the slope and the zero-point, both
depending on $\beta$. The zero-point turns out to depend on $k_v$, $L'_0$ and \Ie. This explain why the \MR\ relation
presents different distributions according to the values of the S\'ersic index, the age of the galaxies \citep{Valentinuzzi2010}
and mean surface brightness \citep[see e.g.][]{Almeida}.  The massive passive galaxies with large negative values of $\beta$
converge towards values of the slope close to 1 (that predicted for virialized systems). On the other hand the lower slope
observed for spiral galaxies is also in good agreement with values of $\beta$ close to $\sim3$.

In conclusion we have seen that the combination of the virial theorem and the FJ-like relation can explain
the observed trends in the \IeRe\ and \MR\ relation, and also the FP \citep{Donofrio2017}.}

\subsection{The Zone of Exclusion (ZoE)}\label{Zoe}

\begin{figure}
    \centering
     {   \includegraphics[width=0.45\textwidth]{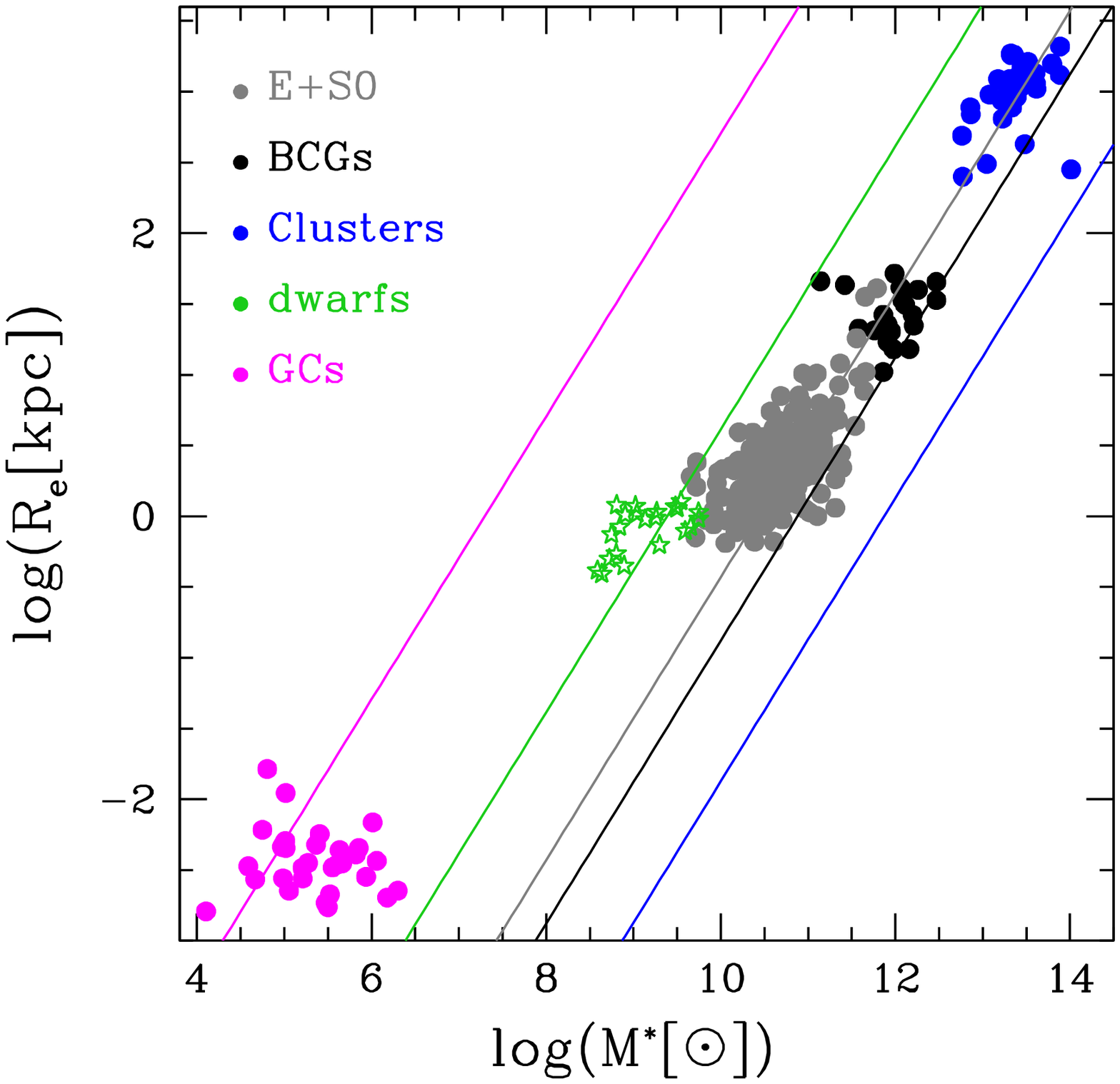}
         \includegraphics[width=0.45\textwidth]{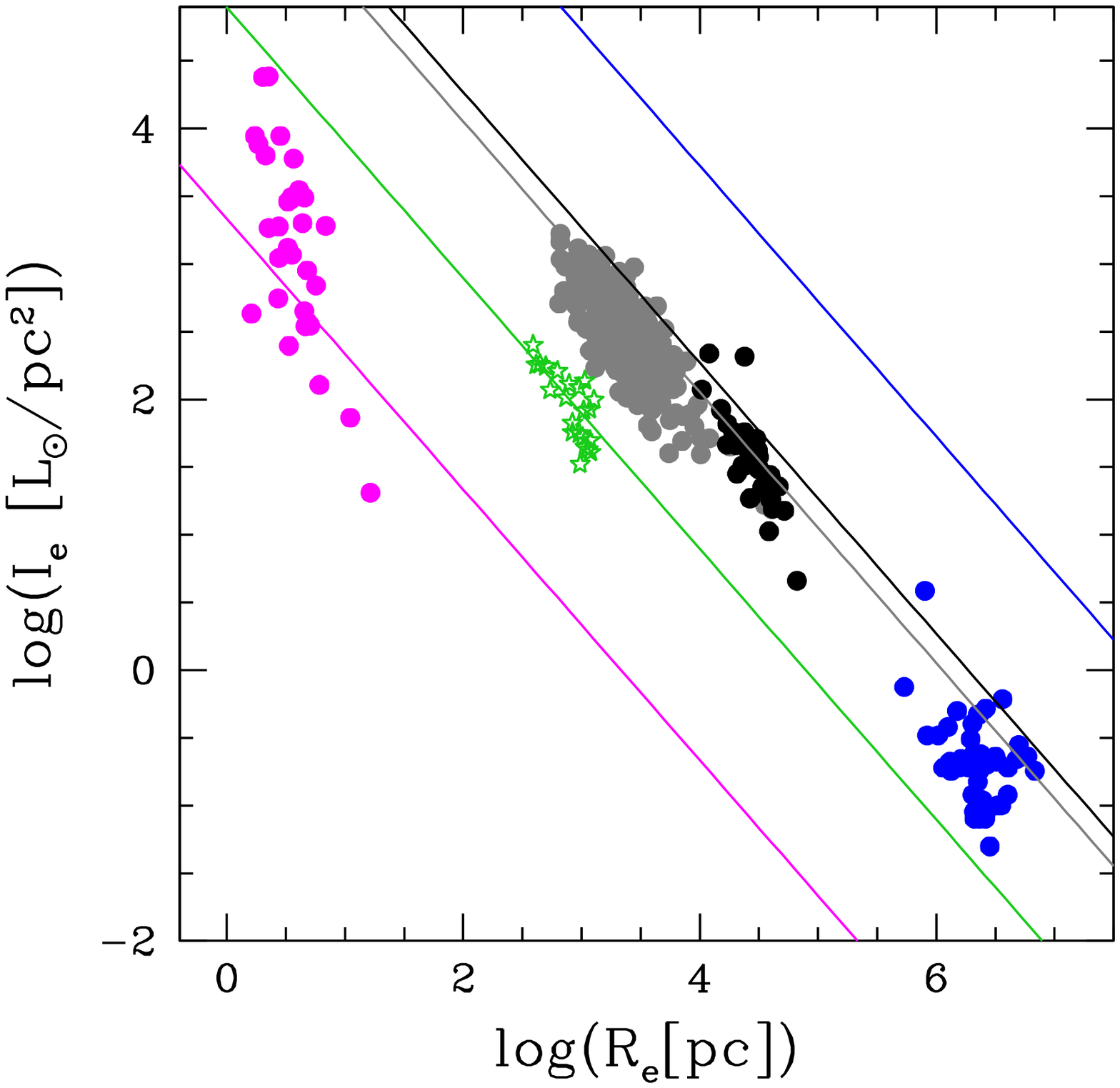} }
    \caption{Upper panel: the \MR\ plane. GCs are marked by magenta dots, dwarfs by green stars, normal ETGs
by gray dots, BCGs by black dots and clusters by blue dots. Each colored line marks the average zero-point of
the systems calculated  from Eqs. \ref{eq1} and \ref{eq6} with the values of $k_v$, $M^*/L$ and $\sigma$. The
solid black line is  obtained for the sample of galaxies and clusters taken together. All the lines have  the
slope 1 predicted for virialized systems. Lower panel: the \IeRe\ plane. The same color code is adopted. Here
the slope of the lines  is $-1$.}
    \label{fig:ZoE}
\end{figure}

Up to now we have suggested that the slope of the ZoE both in the \MR\ and \IeRe\ planes could be that
predicted by the virial theorem
for fully relaxed systems. The slope is $-1$ in the \IeRe\ plane and $1$ in the \MR\ plane. In these figures
we have always drawn the possible ZoE with dashed lines. The problem now is:
What is the zero-point of the ZoE?
This can be derived from Eqs. \ref{eq1} and \ref{eq6} once the values of $k_v$, $M/L$ and $\sigma$ are known.
Unfortunately the total mass $M$ of our systems is unknown, but we can have an idea using $M^*$.
The value of $k_v$ for every system can be approximately obtained from the S\'ersic index $n$ using eq. 11 of \cite{Bertin2002}
(considering only the structural non-homology). The stellar masses of galaxies are known from the SED fitting of the spectra
and from the
stellar mass-to-light ratios of clusters measured by \cite{Cariddietal2018}. The stellar velocity dispersion
is also available for many objects from the WINGS database.
Fig. \ref{fig:ZoE} shows with lines of different colors the different zero-points calculated for our systems
in the \MR\ and \IeRe\ planes. We have added here a sample of Globular Clusters ( the magenta points).
These systems are likely in a good virial equilibrium state and can therefore be used as reference comparison objects.
The data are those of \cite{Pasquato}. For globular clusters we assumed the perfect homology with a S\'ersic index $n=4$.

Figure \ref{fig:ZoE} shows in the upper panel the \MR\ plane with the objects of our sample: Globular Clusters
(magenta dots), normal ETGs (gray dots), faint ETGs (green dots), BCGs (black dots) and clusters (blue dots).
The solid colored lines mark the virial relations (with slope 1) with the different zero-points calculated
for each system: $ZP_{MR}=G/(k_v\sigma^2)$. We take as $ZP_{MR}$ the average value of all zero-points for each
sample of objects considered. Note that the predicted linear trends with these calculated zero-points
intercept the distribution of each sample. The lines however do not cross the distribution exactly in the
middle. This is due to the fact that the variable $\sigma$ depends on the total mass of the system, while here
we are considering the virial relation using the stellar mass. We will see below that using eq. \ref{eq1} we
can get the velocity dispersion $\sigma_*$ that a galaxy would have if DM were absent.

In the bottom panel we can see the \IeRe\ plane where we have calculated the zero-points of our systems
through the formula: $ZP_{IeRe}=(k_vL\sigma^2)/(2\pi GM^*)$. The colored dots mark the same sample of objects.
Again note that the location of the zero-points provide virial lines intercepting each system, but not in the
middle of the observed distribution.

A further thing to note is that the zero-points of systems more massive than $10^{10} M_{\odot}$ are
approximately similar and seem to converge toward the limit of the ZoE. We have checked that the combination
of the variables $k_v$, $\sigma$ and $M^*/L$ is such that very similar values are obtained in log scale for
all these systems.

\begin{figure}
    \centering
	\includegraphics[width=0.45\textwidth]{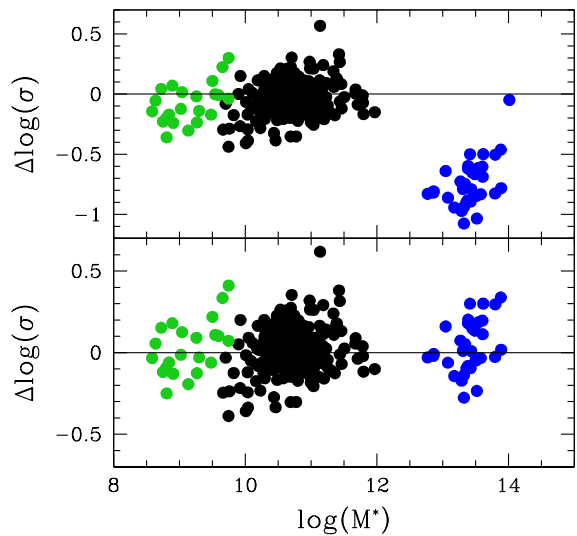}
    \caption{Plot of the stellar mass versus the difference in log units of the velocity dispersion measured
from spectra and calculated through the virial equation. The upper panel shows the difference before the
correction of the stellar mass. The lower panel indicate that once the mass is corrected for the contribution
of DM, the two quantities are in good agreement (see text). Black dots are normal ETGs, green dots the faint
ETGs measured by \cite{Bettoni2016}, blue dots are galaxy clusters.}
    \label{fig:DS}
\end{figure}

In Fig. \ref{fig:DS} we show the difference between the measured $\sigma$ and those calculated from the
virial equation, before (upper panel) and after (bottom panel) a correction applied to the stellar mass $M^*$.
In order to have a mean difference equal to zero we need to correct the stellar masses of the following
quantities:
a factor of 1.66 for dwarfs, 1.25 for normal ETGs and a factor of $\sim40$ for galaxy clusters.
These objects appear indeed dominated by the DM. We have not considered GCs, because they are not affected by DM and
they can loose mass during their crossing of the Milky Way disk.

In conclusion of this section we can say that all our objects are in virial equilibrium. However, as we will see
in the next section, the simulations suggest that the virial equilibrium might be continuously disturbed
by merging, stripping and interaction events. While in massive galaxies the impact of merging, stripping, interactions
with objects of smaller mass can be of minor relevance,  in  dwarf galaxies  these
events may induce severe disturbances. The inner total energy of dwarf galaxies can
significantly vary when even minor mergers occur. In this sense we can introduce the concept of and
speak of a condition of 'undisturbed  virialization' for a galaxy when no more merging/stripping and star
formation events are in
place or when the galaxy is so massive that is no more affected by the small merging or stripping events.

If our view is correct, the ZoE is the natural border of undisturbed  virialized systems that have
reached the maximum possible values for $\sigma$, $k_v$ and $M/L$. We do not know its exact position, being
the DM contribution unknown, so we have chosen an arbitrary value either in Fig. \ref{fig:IeRe} and
\ref{fig:MR}.

Here below by means of simulations we will see that the condition of 'undisturbed  virialization' has been
reached only
by massive ETGs. They are passive and quenched objects. They are so massive that new encounters or mergers do not
significantly alter their virial equilibrium.

\section{Evolution of the scaling relations with redshift}\label{sec:4}

In this section, with the aid of the Illustris library of galaxy models, we have examined the cosmic
evolution of the above seen SRs. We have used the whole dataset of simulated objects with mass larger than
$10^{9} M_{\odot}$ at $z=0$ present in the selected clusters. Each galaxy is followed along its evolutionary
tree (in this case along the "main progenitor branch", \ie\ that following the mass history, see
http://www.illustris-project.org/data/docs/specifications/) since $z=4$, an history made of merging events,
tidal interactions, periods of quiescence, as well as BH and SNe activities.

Prior to any other consideration, by means of SSPs of different
ages and metallicities we examine the luminosity evolution of a galaxy,   either in isolation or in presence
of bursts of star formation that are triggered by mergers with galaxies of comparable mass. Masses are set
equal to those of typical galaxies in the mass interval $10^7$ to $10^{12} \, M_\odot$. We call this type of
galaxy model {\it SSP-galaxies}. Starting from this, we set up  Monte-Carlo simulations of bursts of star
formation in already existing seed objects of given mass, age, and metallicity at varying the burst age and
intensity and/or mergers among galaxies of different mass, age, and metallicity, at varying the the
time of fusion. The details of the Monte-Carlo method
are shortly given in Appendix \ref{Appendix_B}.

In Fig. \ref{fig:Mv_Age}  we show the luminosity evolution of
a $10^9\, M_\odot$ SSP-galaxy undergoing bursts of star formation of different age and intensity
(expressed by the percentage  mass going into stars). In our simulations this percentage is assumed  equal
to 30\% of the galaxy mass. The age of the galaxy is fixed to 13 Gyr and its mean metallicity is estimated
from the mass-metallicity relation of Table \ref{mass_metals} given in Appendix \ref{Appendix_B} and taken
from \citet{Sciarratta_etal_2019}. The ages of the bursts are  7 Gyr (oldest), 2 Gyr, and 1 Gyr (youngest).
As expected the
luminosity evolution expressed by the absolute visual magnitude $M_v$, depends on the burst age. The oldest
one is in practice indistinguishable from the case of the unperturbed galaxy (thick black line), whereas for
the youngest one, we expect a present day absolute magnitude about 1 mag brighter than the unperturbed case.

Similar results are obviously possible at varying the galaxy mass. This is achieved by  simply scaling up and
down by the luminosity of the  $10^9 M_\odot$ objects by the
quantity $10^{-0.4\Delta M}$.

\begin{figure}
	\centering
	\includegraphics[width=0.45\textwidth]{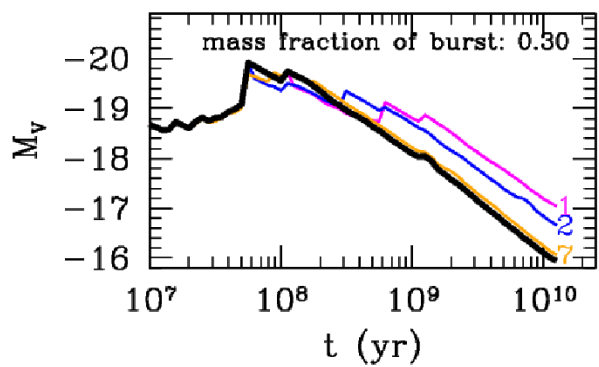}
	\caption{Results of dissection of single bursts inside 13 Gyr-old SSP-galaxies:
		evolution with time of $M_V$ for the SSP-galaxy
		with $\log M=9$ ($Z=0.004$) perturbed by
		a single burst with mass fraction of $30\%$; solid thin curves are, in terms of increasing age,
		magenta (1 Gyr), blue (2 Gyr)
		and orange (7 Gyr); the thick black line is the unperturbed case shown for comparison.    }
	\label{fig:Mv_Age}
\end{figure}

Mergers among galaxies of different mass and age would yield
similar results, the variation in absolute magnitude (luminosity) being driven by the variation
in mass and  age of the two component galaxies, together with a small  contribution due to different mean
metallicity of the galaxies. If mergers are also accompanied by revival of the star
formation activity an additional variation in the present day luminosity is expected. Finally if a galaxy
of a certain age and mass suddenly stops star formation, its luminosity would soon fall onto the
luminosity-age relation ship of the passive case (the time scale involved would be of the order of 1 Gyr or
less).

The main conclusion is that the luminosity (in the V pass-band in this case) significantly depends on the
particular star formation history of each galaxy, in such a way that it cannot be easily traced back from
the present day properties. This example clearly shows that a
significant dispersion in the V-luminosity of a galaxy of given mass is possible and also
expected. This would   blur the proportionality factors $L_0$ and/or $L'_0$ of eqs. \ref{eq2} and/or
\ref{L_sig}. The expected blurring in luminosity is  $\Delta\log(L) \simeq 0.4$, \ie\ very close
to the observed dispersion in the \Lsig\ relation.

\begin{figure}
    \centering
	\includegraphics[width=0.45\textwidth]{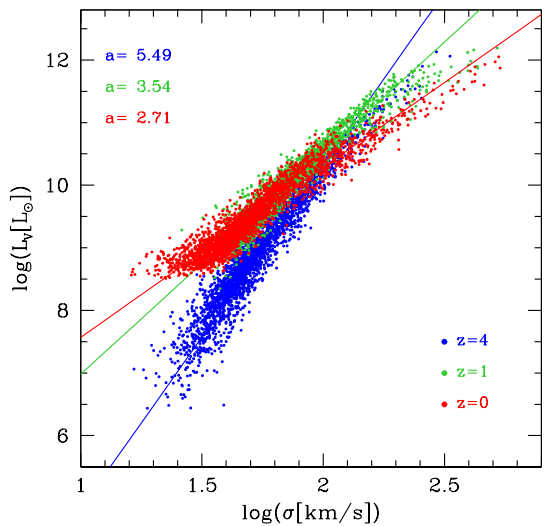}
    \caption{Simulations of the \Lsig\ relation with the galaxies of the Illustris dataset at three different
redshifts. Objects at $z=0$ are marked by red dots, at $z=1$ by green dots, and at $z=4$ by blue dots. The
colored lines show the best fits of the distributions and the
resulting slopes are listed in the top left corner.}
    \label{fig:sim_logL_sigs}
\end{figure}

Fig. \ref{fig:sim_logL_sigs} shows the \Lsig\ relation expected from the Illustris simulation at different
cosmic epochs:  galaxies at $z=4$ are marked by blue dots, at $z=1$ by green dots and at $z=0$ by red dots.
It is worth recalling that the objects at $z=4$ are the progenitors of those at $z=1$ and the latter in turn of
those at $z=0$.
We clearly see that going toward the present epoch the global distribution of galaxies is progressively
less steep, but the scatter
is very similar. The slope/rms decreases from 5.49/0.24 at $z=4$ to 3.54/0.17 at $z=1$ and to
2.71/0.18 at $z=0$.
Note that the \Lsig\ relation is rather narrow at any redshift. A little change in slope seems also be
present for the brightest galaxies after $z\sim1$, with a smooth flattening of the
relation toward lower slopes.

Globally the relation seems to rotate with time around a point approximately located at $\sigma=100$
km s$^{-1}$ and $L=10^{10}$ $L_{\odot}$.
This means that on average the points move in a direction almost perpendicular to the observed relation
reinforcing the idea presented
in Sec. \ref{sec:2} that the \Lsig\ relation should be written in the form of eq. \ref{eq5}, where
both $L'_0$ and $\beta$ are variables. In this way galaxies can move in this plane along paths that depend on
the peculiar merging/interaction events and on the SFH.

Finally, there is another important remark to be made, looking at the \Lsig\ relation at different redshifts. In
Fig. \ref{fig:sim_logL_sigs} we note that {at a given redshift the dispersion $\Delta log L_V$ decreases
at increasing $\sigma$ (mass) of the galaxy. The explanation relays on the merger mechanism itself: by
increasing the mass of a galaxy the probability of merging another object of comparable mass (so that the
effect on the luminosity would be sizable) decreases in compliance to the number density at varying the
galaxy mass, the so-called funneling effect amply discussed by
\citet[][references therein]{Sciarratta_etal_2019}. As a result of it, as the redshift tends to zero, high
mass objects engulf galaxies of small mass so that the net effect on their luminosity becomes very small. }

The evolution of the \Lsig\ relation is in line with the moderate evolution of the FP coefficients found
by \cite{Luetal2019} from $z=0$ to $z=2$ with
the data of the IllustrisTNG simulation {(Alberini et al. 2020, work in progress)}.

\subsection{The possible values of $\beta$}

{For what we have shown before the values of $\beta$ for each galaxy are of big importance to understand the
global behavior of the SRs. The galaxies move  in the \IeRe\ and \MR\ plane in a direction fixed by the values
of $beta$. Here therefore we try  to estimate the values of this parameter using the results of simulations. The
starting point is to recognize that $beta$ is given by the slope of the line connecting two different location of
the same galaxy in this plane.}

\begin{figure*}
	\centering
	\includegraphics[width=1\textwidth]{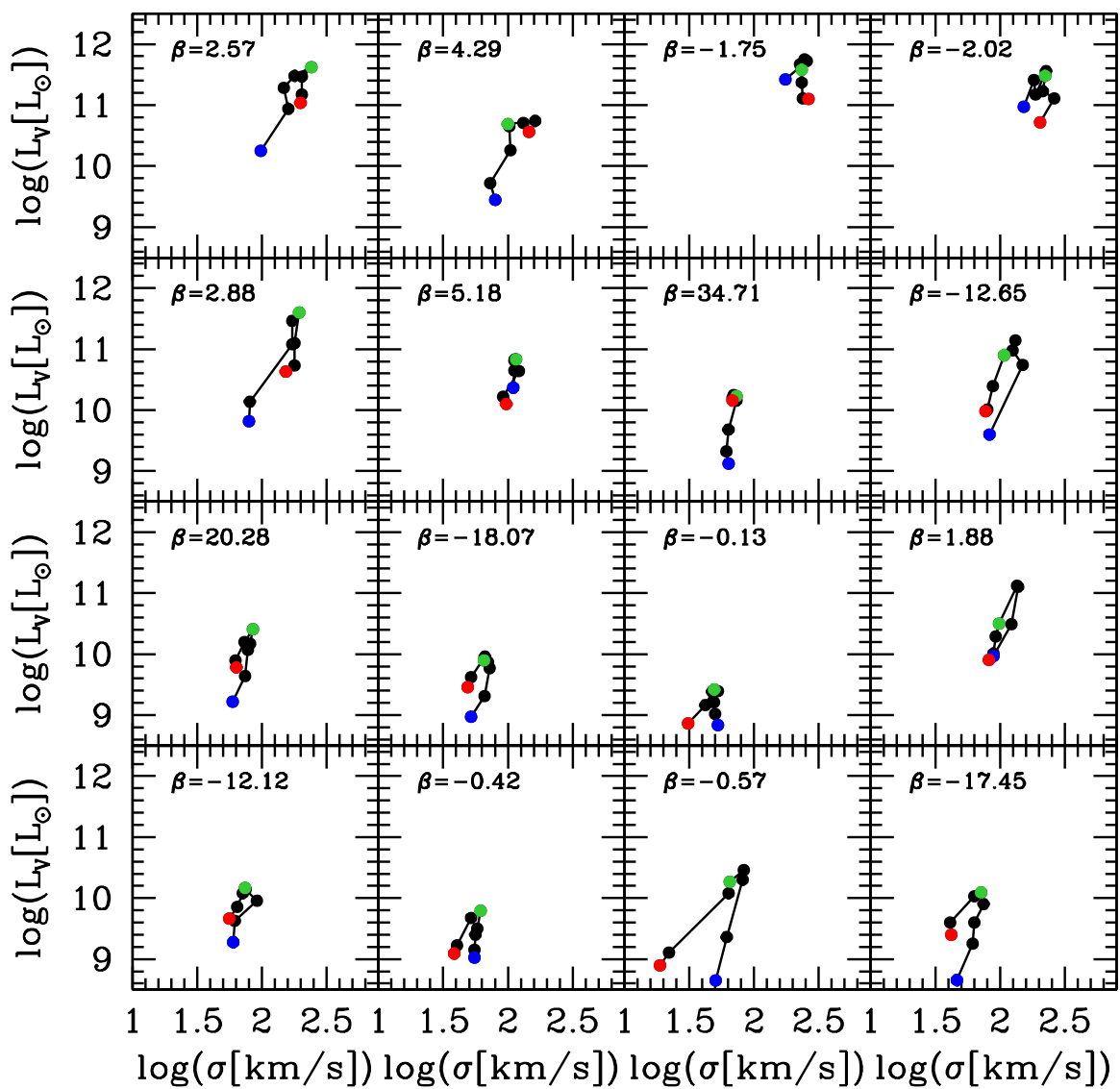}
	\caption{The path of single galaxies in the \Lsig\ plane from $z=4$ (blue dot) to $z=0$ (red dot). Each box
		list on the top the value of the slope $\beta$ of the trajectory connecting the two epochs.}
	\label{fig:paths}
\end{figure*}

In Fig. \ref{fig:paths} we see the paths of few single galaxies in the \Lsig\ plane from $z=4$ to the present.
They are complex and clearly mirror the effects of several variables. Each path is made of many steps in
which the mass and velocity dispersion are varied. In general there are long steps in which the mass is
significantly increased/decreased by mergers/interactions, and short steps in which the mass and velocity
dispersion vary by small amounts. The steps may have different inclinations in the \Lsig\ plane.
The evolution starts at $z=4$ (blue dots) and goes through $z=1$ (green dots), ending at $z=0$ (red dots).
The black lines follow each path along the various redshift epochs.

\begin{table}
\begin{center}
\caption{Average and median values of $\beta$ for the two intervals in cosmic epochs in the different mass ranges.}
		\label{Tab_beta_mass}
		\begin{tabular}{|c| c| c|}
\multicolumn{3}{c}{ From $z=4$ to $z=0$} \\
\hline
         $M^*$ range    &  mean $\langle\beta\rangle$  & median $\Tilde{\beta}$  \\
\hline
         $M^*>10^{11}$     &        120.2   &      2.2 \\
         $10^{10.5}< M^* \leq10^{11}$      &       3.4     &       4.6   \\
         $10^{10.0}< M^* \leq10^{10.5}$      &     8.5     &       6.5   \\
         $10^{9.5}< M^* \leq10^{10.0}$      &    -61.9     &       5.5   \\
         $10^{8.0}< M^* \leq10^{9.5}$      &       11.0     &       0.9   \\
\hline
\multicolumn{3}{c}{ From $z=0.2$ to $z=0$} \\
\hline
         $M^*$ range    &  mean $\langle\beta\rangle$  & median $\Tilde{\beta}$  \\
\hline
         $M^*>10^{11}$     &       -0.7   &      2.1 \\
         $10^{10.5}< M^* \leq10^{11}$      &       -9.1     &       4.5   \\
         $10^{10.0}< M^* \leq10^{10.5}$      &     -2.5     &       3.4   \\
         $10^{9.5}< M^* \leq10^{10.0}$      &      4.3     &       3.7   \\
         $10^{8.0}< M^* \leq10^{9.5}$      &       5.7     &       3.1   \\
\hline
\end{tabular}
\end{center}
\end{table}

\begin{table}
\begin{center}
\caption{The different values of $\beta$ and the corresponding slopes in the \IeRe\  and \muere\ planes.}
		\label{beta_values}
		\begin{tabular}{|r| c| c| c|}
\hline
         $\beta$     &  Ie-Re  & mue-Re & R-M*  \\
\hline
         3.0     &        1.0   &      -2.50 & 0.43 \\
         2.0     &        -      &        -      & 0.33 \\
         1.0     &      -3.00   &      7.50  & 0.20 \\
         0.5     &      -2.33   &      5.83  & 0.11 \\
        -0.5     &      -1.80   &     4.50  & -0.14 \\
        -1.0     &      -1.66   &     4.16  & -0.33 \\
        -1.5     &      -1.57   &     3.92  & -0.66 \\
        -2.0     &      -1.50   &     3.75  & -1.00 \\
        -2.5     &      -1.44   &     3.16  & -1.66 \\
        -3.0     &      -1.40   &     3.50  & -3.00 \\
        -3.5     &      -1.36   &     3.41  & -7.00 \\
        -4.0     &      -1.33   &     3.33  &  0.00 \\
        -4.5     &      -1.30   &     3.26  &  9.00 \\
        -5.0     &      -1.28   &     3.21  &  5.00 \\
        -8.0     &      -1.20   &     3.00  &  0.50\\
       -11.0    &      -1.15   &     2.88  &  1.57 \\
       -25.0    &      -1.07   &     2.68  &  1.19 \\
       -50.0    &      -1.03   &     2.59  &  1.08 \\
      -100.0   &      -1.01   &     2.55  &  1.04 \\
     -1000.0  &      -1.00   &     2.50  &  1.00 \\
    -10000.0 &      -1.00   &     2.50  &  1.00 \\
\hline
\end{tabular}
\end{center}
\end{table}

{Empirically we can define a 'mean path', lets say from $z=4$ to $z=0$, considering the line connecting the two points
(blue and red) in this diagram and an 'instant path', connecting the two points at redshift $z=0.2$ and $z=0$
(the two closer epochs). The exact value of $\beta$ today is unknown. We can only estimate its values during past intervals
of time that have seen a galaxy to change its luminosity and velocity dispersion.}

On the top left of each panel in Fig. \ref{fig:paths} we have listed the value of $\beta$, the exponent
that enters in the \Lsig\ relation that can be
obtained measuring the slope of the line connecting the points at $z=4$ and $z=0$.
Note the high spread of values of $\beta$, spanning either negative and positive values.
Positive slopes up to about 5 are expected
in presence of mergers among galaxies of comparable mass. Higher positive values deserve some care and
attention because mergers among galaxies of similar mass are becoming less important and other secondary
 effects on the \Lsig\ relation
could  show up. Very high negative slopes (say below -5) are also of interest because they indicate the
presence of important episodes of mass removal (thus masking the effect of the initial redshift on the
velocity dispersion).  Particularly interesting are the cases with  negative slopes in the bin 0 to $-5$,
which are very frequent (this is indeed the second populated bin of the distribution in the domain of negative
slopes) and the mean slope of the whole sample with redshift from $z=4$ to $z=0$ which is close
to $-1$. Finally, very negative $\beta$ are those belonging to passive systems; quenched objects were the luminosity
if progressively decreasing.

\begin{figure*}
    \centering
     {   \includegraphics[width=0.45\textwidth]{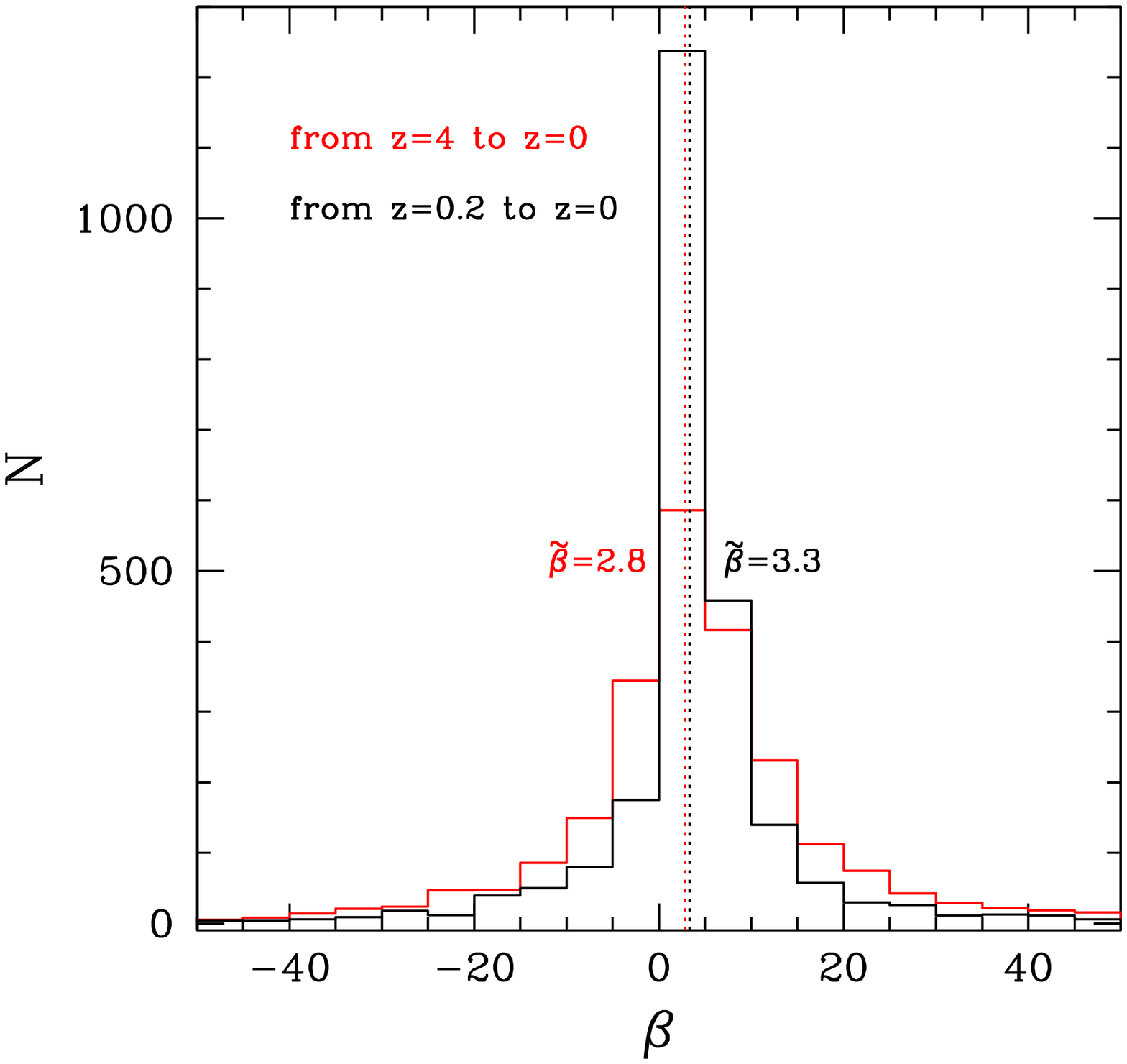}
         \includegraphics[width=0.45\textwidth]{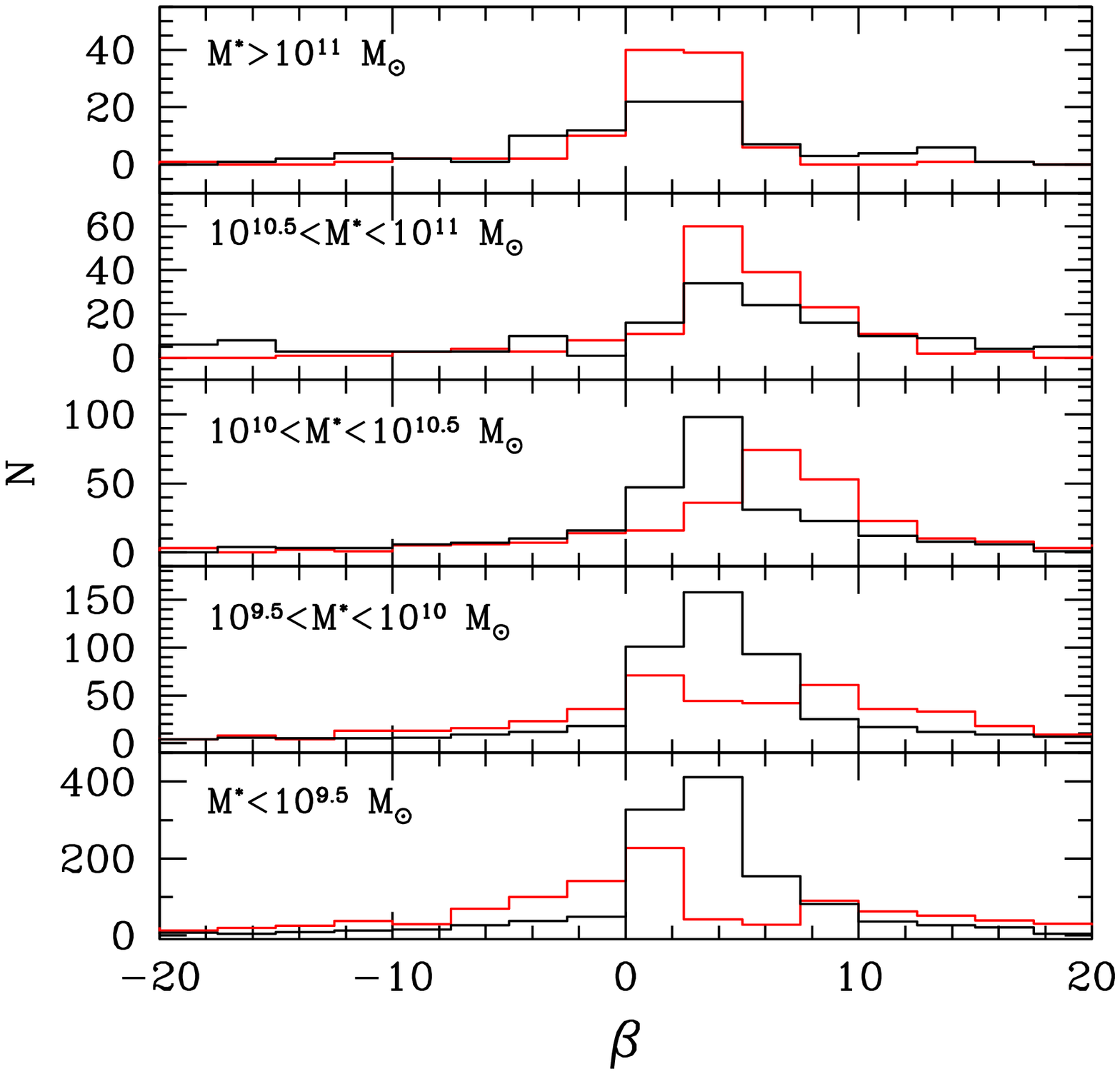} }
    \caption{Left panel: The distribution of the slope $\beta$ for the whole sample of galaxies.
The red histogram is that connected with the values of $\beta$ measured from $z=4$ to $z=1$.
The black histogram that for $z=0.2$ to $z=0$. The dashed lines mark the medians of the distributions.
Right panel: The distribution of the slope $\beta$ for the galaxies of different masses.
The red histogram is that connected with the values of $\beta$ measured from $z=4$ to $z=0$.
The black histogram that from $z=0.2$ to $z=0$.}
    \label{fig:mean_slope}
\end{figure*}

We have evaluated the mean slope of the paths for two different groups in redshift, i.e. from $z=4$ to
$z=0$ (galaxies followed up to the far past \ie\ $\sim$12.1 Gyr ago) and from $z=0.2$ to $z=0$
(galaxies followed up to the recent past \ie\ $\sim$2.4 Gyr ago).
The resulting slopes for the two intervals in cosmic epochs are shown in the left panel of Fig. \ref{fig:mean_slope}.
The two distributions of $\beta$ peak in the interval 0 to $\simeq 3 \div 4$ and nearly symmetrically
extend to very high negative and positive slopes. The average slope is $-1$ for the case in which galaxies are
followed from $z=4$ to $z=0$, while it is $\sim3$ for the case containing galaxies traced back from $z=0.2$ to $z=0$.
On the other hand the medians both peak around $\sim3$.
The red histogram shows the values of $\beta$ measured for the lines connecting the dot at $z=4$ with
the dot at $z=0$. The black one instead gives the distribution of $\beta$ for the more recent epoch
(from $z=0.2$ to $z=0$). The median values of the two distribution are reported in the plot. Notably the
median values peak approximately at the slope observed for the real \Lsig\ relation. This means that the fit
of the observed distribution is primarily influenced by the complex history of mass assembly of the single
galaxies. Note that the most common path corresponds to the slope of the observed FJ relation.

The right panel of Fig. \ref{fig:mean_slope} shows the same histograms for different bins of galaxy masses.
The average slope varies considerably for the different mass ranges (see Table \ref{Tab_beta_mass}), while the
median is always positive. This implies that galaxies of different masses experience different events with
different consequences.

If we differentiate eq. \ref{sig_mass_zeta} and \ref{L_sig_zeta} given in Appendix \ref{Appendix_A}, we can
get an idea of the main contributions in $\Delta L$ and $\Delta\sigma$ that determine the shifts of the points in the
\Lsig\ plane. The mass term dominates, while the other terms do not contribute in a significant way.

\begin{figure*}
    \centering
\includegraphics[scale=0.8,angle=0]{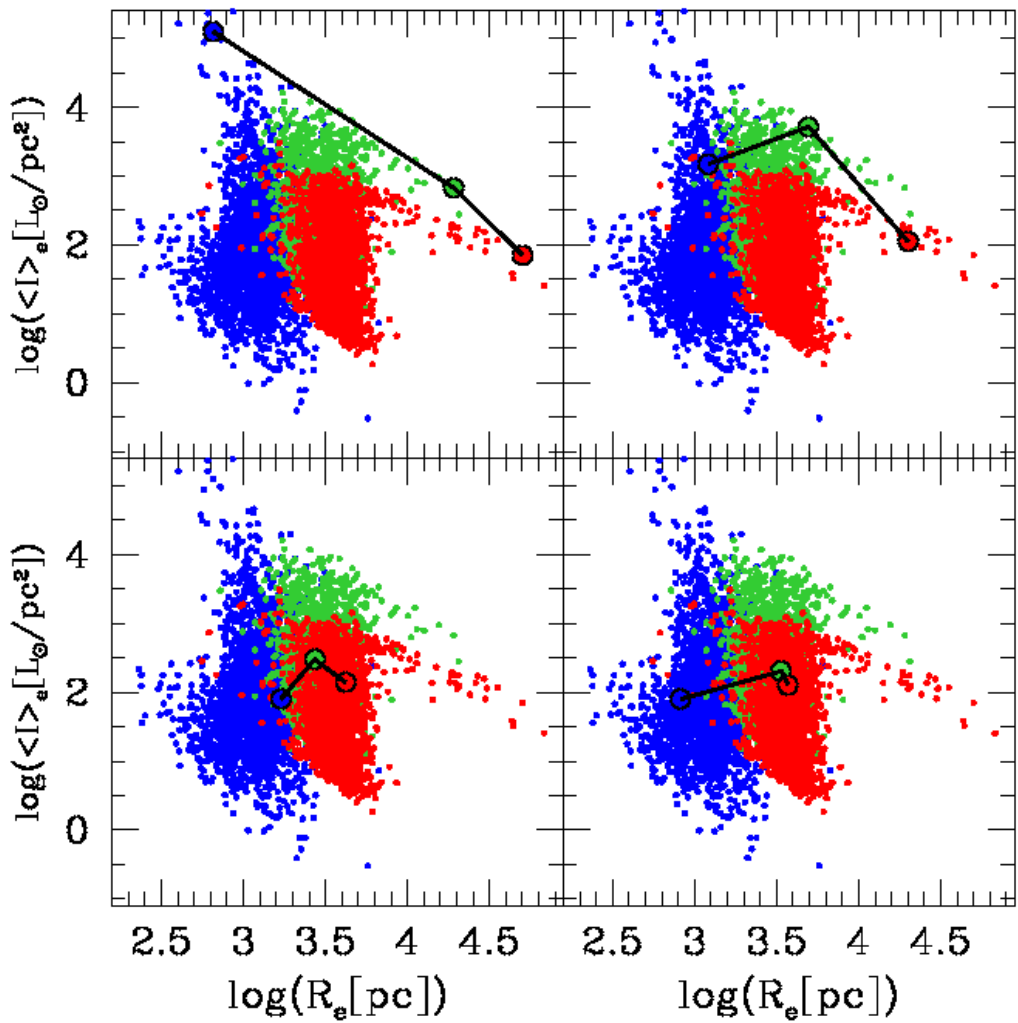}
    \caption{Four different paths in the \IeRe\ plane resulting from simulated data. Blue dots mark the
distribution at $z=4$, green at $z=1$ and red at $z=0$. The half-mass radius in pc unit has been assumed to be equal to
the effective radius. The black lines connect the same object at different epochs.}
    \label{fig:IeReS}
\end{figure*}

Fig. \ref{fig:IeReS} shows the paths of four galaxies in the \IeRe\ plane. In the figure we have marked in
blue the galaxy distribution at $z=4$, in green that at $z=1$ and in red that at $z=0$. In the upper panels
the black lines show the evolution of two galaxies that at $z=0$ are observed in the tail of the \IeRe\
relation (\ie\ objects belonging to the 'bright' family), while in the bottom panels that of objects of the
'ordinary' family. In general the paths are very different for each galaxy: the position in the diagram
appears strongly influenced by the mass assembly history.

Note that large positive values of $\beta$ produce positive slopes in the \IeRe\ plane that could not belong
to objects of the bright family. These objects have already reached a passive evolution. Positive values of
$\beta$ can be observed only for galaxies of the 'ordinary' family. {On the other hand large negative values of
$\beta$ converge toward a limiting slope in both the \IeRe\ and \MR\ relation (see table \ref{beta_values}).}

The formation of the 'bright' family tails in the \IeRe\ and \MR\ planes is very interesting. The simulations
 are in fact able to reproduce such peculiar features: the observed distributions of bright galaxies in the
\IeRe\ plane and the steeper part of the \MR\ relation. Both sequences are formed by objects with mass larger
than $10^{10} M_{\odot}$. How they originate? We have seen from simulations that these tails are absent at
earlier epochs (before $z=2$).
If the tails originate from the merging activity, what kind of merger is it? We have
speculated that dry mergers should be responsible of these features. The merging of stars without gas might
in fact inflate the systems, because the global energy is not dissipated by heating the gas. The absence of
gas is also apparent from the fact that there are not star formation associated (the tails are made by the
most red galaxies).

\begin{figure*}
    \centering
\includegraphics[scale=0.8,angle=0]{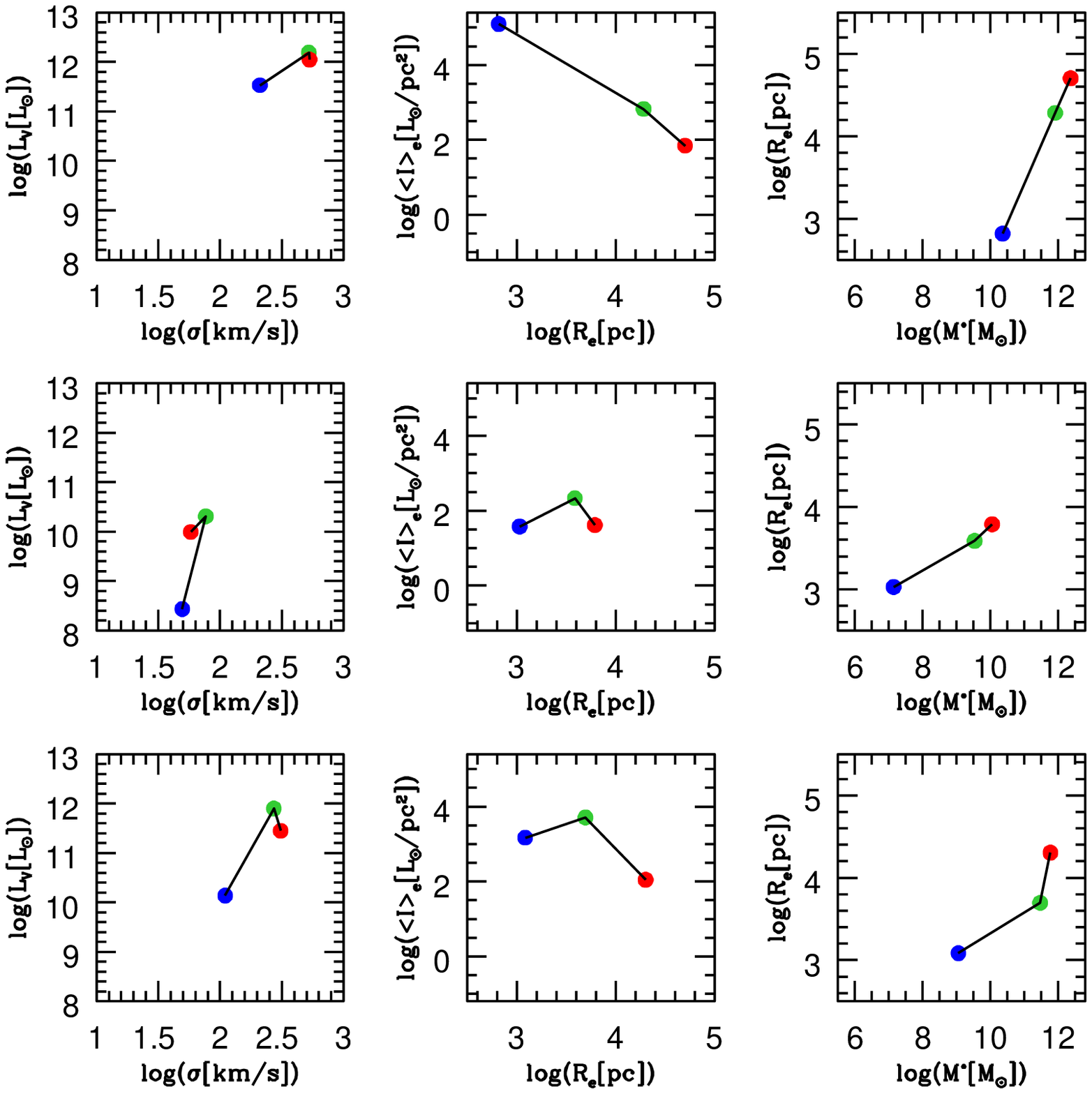}
    \caption{The paths of three ETGs in the \Lsig, \IeRe\ and \MR\ planes. Blue dots mark the position at
redshift $z=4$, green dots that at $z=1$ and red dots that at $z=0$. }
    \label{fig:Allpaths}
\end{figure*}

Fig. \ref{fig:Allpaths} shows the paths of three galaxies in the \Lsig\ (left panel), \IeRe\ (middle panel)
and \MR\ (right panel) planes. Again dots of different colors
mark the position at different redshifts. Note that the ETGs that in the \MR\ plane have the largest mass and
radius, in the \Lsig\ plane move toward a lower luminosity (\ie\ have a negative slope $\beta$) and in the
\IeRe\ plane belong to the 'bright' family. In the middle panels we can see the path of an object that
does not belong to the tails is a member of the 'ordinary' family.
The simulations seem to indicate that a positive variation in mass is not always accompanied by a positive
variation in radius and luminosity.

What appears to originate the observed tails, that we have identified as the SRs, seems more connected with
the existence of the ZoE. When a galaxy reach the passive state can also fully relax and become virialized.
The ZoE could therefore be a sort of universal limit established by the condition of full
virialization and passiveness.
The ZoE indicates that an object of a given mass can never have a radius smaller than that achieved when
it reach the undisturbed virialization and passive state.
Since no system can cross the ZoE, this line appears as the physical driver of the \IeRe\ and \MR\ SRs. Only
the systems that have reached a full virialization and are today evolving in a pure passive way could be
distributed along the tails. The virial SRs with similar zero-points seem to appear only when these conditions
are met. This occurs for the massive galaxies that are today poorly affected by minor mergers (major mergers
are very rare), so they are the systems closest to the condition of full virialization. They are also passive
objects since their star formation quenched long time ago. For the objects of the 'ordinary' family the virial
equilibrium is very unstable, since merging and stripping events and episodes of star formation rapidly move
the galaxies toward a new condition of virial equilibrium. These systems are not passive yet and are therefore
far from the ZoE.

In paper III we will address the question of the ZoE more deeply, examining the possible role played by cosmology.

{Finally we remark that simulations show the formation of these tails only for galaxies with redshift $z<2$.
The tails are well visible at $z=0$ only for massive systems. Both
the \MR\ and \IeRe\ tails do not exist before that epoch. We argue that the origin of these tails is the same
for both planes. It is due to the progressive variation of homology of massive systems caused by the large number of
dry merging events. These galaxies are almost passive and have developed a large extended stellar halo.
Their S\'ersic index is big, so that the combination of $k_v$, $\sigma$ and $M^*/L$ in log scale is
progressively converging toward the limit of the ZoE. On the other hand, the small galaxies follow an almost
flat distributions in these planes at any redshift. The typical values of $\beta$ for these systems is $\sim 3$;
they are objects moving along the slope of the observed FJ relation. Their dynamical status is continuously changed
by interactions and feedback effects.}

\section{Conclusions} \label{sec:5}

By exploiting the data of the WINGS and Omega-WINGS surveys we have investigated the distribution of galaxies and GCs
in the \Lsig, \IeRe\ and \MR\ planes. Then using the data extracted from the "Illustris"
simulation, we have compared the SRs resulting from the hydro-dynamical models with the observational ones.
In summary these are our main conclusions:

\noindent (-) Galaxy clusters follow the same SRs of BCGs: their location in the \Lsig, \IeRe\ and \MR\
planes is that of very large, bright and high velocity dispersion BCGs.
In paper I we noted that the normalized light profiles of galaxy clusters can be superposed to that of normal
ETGs of intermediate luminosity. In this case therefore the parallelism with ETGs is
with the brightest systems and not with the less luminous objects. From the equivalence of the normalized
profiles one can argue that the density distribution of galaxies in clusters is in some way similar to that of
galaxies of intermediate/faint luminosity. Their structural parameters on the other hand are those of very
bright and big BCGs. How can we explain this behaviour?
A possible answer is that the original mass profile of all these systems were approximately the same at earlier epochs
(as we suggested in paper I), but BCGs have progressively modified their profiles for the modifications induced by
feedback effects and merging events. These modifications have not affected the GCs
considered in our study. They are likely systems close to the virial equilibrium, with
light profiles well fitted by a single S\'ersic law. There are many nearby clusters ($\sim30\%$) still far
from this condition, that do not follow the same SRs of virialized clusters
\citep[see][]{Cariddietal2018}.
The transformation of the inner and outer density distribution of BCGs has probably no significant effects
on the effective radius of these galaxies. This might occur if the mass fraction involved in the transformation is
low in comparison with the total mass of the system. In Fig. 4 of paper I we can see that
the light profiles of faint and bright ETGs differ in the ranges $r<0.15 R_e$ and $r>2.5 R_e$, \ie\ in the zones
including a small fraction of the total mass.
The bulk of the mass (and consequently of the light )is contained in the interval $0.15<r/R_e<2.5$. The
size of the effective radius depends on the bulk of the mass assembly and not on the mass involved in the transformation.

\noindent (-) The numerical simulations reproduce quite well the distribution of the BCGs and II-BCGs in the
\Lsig, \IeRe\ and \MR\ planes, while seem to fail for dwarfs and galaxy clusters.
The effective radius deduced from the effective mass radius can be still a factor of $\sim3$ larger than
observed for dwarfs. Simulated clusters are in general fainter and  smaller in radius than
real clusters. However, the well known trends visible in the \IeRe\ and \MR\ planes made by bright galaxies
are well reproduced. These relations appear as tails emerging from the flat distribution of less luminous
galaxies. They appear after $z\sim2$, that is after the epoch of maximum star formation, when systems
progressively quenched. The real galaxies show that these trends are better visible for galaxies in clusters,
than for objects in the field.

\noindent (-) The simulations indicate that each galaxy follows a complex path of evolution in the \Lsig,
\IeRe\ and \MR\ planes. This path is due to the chaotic mass assembly history,
made of merging, interaction/stripping events, vigorous star formation and feedback effects. The most
frequent paths determine the mean distribution observed in these planes.
This behavior justify the assumption of writing the \Lsig\ relation in a new form, independent from the
virial theorem:
$L=L'_0\sigma^{\beta}$ law, where the slope $\beta$ can assume either positive and negative values and
$L'_0$ is the key
variable connected to the mass assembly and star formation history. {The values of $\beta$ fully constrain the slopes
(\ie\ the direction of motion) of galaxies in the SRs.} Large negative values of $\beta$ are
those belonging to passive systems, that naturally evolve toward progressively low
values of the total luminosity. Most of the objects that are today in the tails of the distributions
observed in the \IeRe\ and \MR\ planes have $\beta<0$, \ie\ are quenched passive systems. One should keep in mind that
this relation is valid for a single galaxy.

\noindent (-) Both real and simulated data seem to show that there is a ZoE in the \IeRe\ and \MR\ planes,
that is a region forbidden to any type of objects. The origin of this
empty region is not fully understood. No system can cross the ZoE. The slope and zero-point of this line
is the same for any kind of object, independently of its mass.
We have identified this line with the locus of fully virialized and passive objects.
Galaxies progressively grow in mass and size across the cosmic epochs. After $z\sim2$, going toward the
present, we observe in the simulations the formation of two tails in the \IeRe\ and \MR\ planes that indicate
the existence of the ZoE. The most massive galaxies are the oldest virialized passive systems, so they are
distributed almost along the ZoE. We have attributed to dry merging the growth in mass and size of these
systems. Since the galaxies that follow the trend of the ZoE are in virial equilibrium, we argue that the
dry merging events affecting these galaxies should involve small amount of mass that do not alter
significantly the dynamical structure of the galaxies, but only the outer regions.

\noindent (-) Dwarfs galaxies are not distributed along the ZoE; their effective radius does not scale
linearly with the total stellar mass. This could be due to several reasons: the progressive large influence of
DM, the effects of rotation and finally the possibility that these systems are in a pseudo-virial condition,
subject to transformation as soon as new mergers occur.
Possibly many of them have not reached yet the condition of virial passive evolution, so that their radius
could be larger than expected on the basis of the full virialization.
We know that in many of them star formation and galactic winds are still ongoing and many suffer strong
interactions and merging with other galaxies of comparable mass in the clusters that might severely affect
their dynamical equilibrium. Energy is continuously injected in these systems determining a larger radius. In
conclusion, the zero-point of the virial relation seems quite different for any object.
We speculate that the systems that are still growing today, will finally enter in the
tails of the bright virialized objects, while the dwarfs that will not grow anymore once the merging
events will be rare, progressively will settle along the ZoE decreasing their radius. The simulations indicate
in fact that all systems evolve toward the virialization and passiveness.

\noindent (-) The distributions of objects observed in the \Lsig, \IeRe\ and \MR\ planes are connected
each other. The origin of the deviation from the trend predicted by the virial theorem is the same for all of
them. Ultimately it is due to a progressive variation of the mass-to-light ratio and homology induced by the
large number of merging events experienced by galaxies. The small scatter suggest that a fine-tuning between
structure and stellar population is in place.

Notably in the \IeRe\ and \MR\ planes it is well visible
the presence of the ZoE, that does not appear in the \Lsig\ plane. The simulations however suggest a little
change in the slope of the relation for the brightest galaxies, \ie\ for those entering in the tails of the
\IeRe\ and \MR\ relations. What is surprising is that despite the chaotic paths of evolution, the \Lsig\ relation
appears narrow at any epoch.
The reason for this might reside in the moderate luminosity decrease with time of old stellar systems (the
short steps in the total paths on the \Lsig\ plane at nearly
constant mass during which age effects can be seen). In addition to it, the sudden acquisition/loss of mass
by mergers can change the mass and the stellar velocity dispersion relocating a galaxy in a different
position along the
plane (a  merger of two galaxies with equal mass and
luminosity generates an object two times brighter and more massive, i.e.  $\sim 0.3$ dex on both coordinates
in the \Lsig\ plane). This is almost equivalent to the spread observed
in the relation.
In other words it appears that the mass acquisition/loss acts like a ''planer`` ultimately shaping the
\Lsig\ distribution. The mass  is the more important parameter  determining the final luminosity and
velocity dispersion of a galaxy.

Finally we want to point out that in the hierarchical scenario of galaxy formation and evolution, mergers and
interactions drive the structural properties of the galaxies, whereas the natural aging of the stellar
populations plays a concomitant less relevant role.
This role is much evident at the present epoch, when mergers are rare and a passive luminosity evolution takes place.
On the other hand, in the early hierarchical or quasi monolithic view of galaxy formation, mass and velocity
dispersion are acquired very
soon and remain (nearly) constant  ever since, so that only the luminosity changes.
For this reason we expect that the observed distribution of real galaxies in the \Lsig, \IeRe\ and \MR\
planes at increasing redshift will provide in the next future important information on the dynamical process
of mass aggregation, structure formation and evolution of the stellar population, as well as on the importance
of feedback effects.

\begin{acknowledgements}
We like to thanks Prof. L. Secco and Dr. R. Caimmi for very useful clarifications about the Clausius'
Virial Theorem for multi-components systems.
Finally, C.C. thanks the Department of Physics and Astronomy of the Padua University for the hospitality and
computational support.
\end{acknowledgements}

\begin{appendix}
\noindent

\section{Two components Virial Theorem and Extended \Lsig\ Relation} \label{Appendix_A}
The virial equations and the virial theorem
for a composite system made of DM and BM  can be derived for each component in presence of the tidal
potential exerted by the other. We limit ourselves to the simplest case of two homogeneous concentric
spherical subsystems made of DM and BM with isotropic stress tensor. The DM component has mass $M_{DM}$ and
radius $R_{DM}$  The baryonic mass is supposed to be made of stars (and gas that is neglected here), with
total mass $M^*$ and radius $R_s$.  Finally, the baryonic component
is internal to the one made of dark matter.

\citet{Secco_Caimmi_1992,Caimmi2003,Caimmi2007,Caimmi2009} developed the virial equations for each
subsystem taking into account the tidal
potential exerted by the other component. We will strictly follow their formalism.  We start from the
kinetic energies of the two components:

\begin{eqnarray}
2 (E_{BM})_{k} &=& \frac{3}{5} \frac{GM_{BM}^2}{R_{BM}} + \frac{3}{5} \frac{GM_{BM}^2}{R_{BM}}
\frac{x}{y^3}\\
2 (E_{DM})_{k} &=& \frac{3}{5} \frac{GM_{DM}^2}{R_{DM}} + \frac{3}{5} \frac{GM_{DM}^2}{R_{DM}}\frac{x}{y^3}
\left(\frac{5}{2}y^2 -\frac{3}{2}\right)
\label{virial}
\end{eqnarray}
where
\begin{equation}
x= \frac{M_{DM}}{M_{BM}}\qquad  y= \frac{R_{DM}}{R_{BM}} \qquad  y \geq 1
\end{equation}
and $k$ stands for kinetic. In terms of the mass-weighted velocity dispersion, one obtains:

\begin{eqnarray}
\sigma_{BM}^2 &=& \frac{3}{5} \frac{GM_{BM}}{R_{BM}}  \left(1 + \frac{x}{y^3} \right) \\
\sigma_{DM}^2 &=& \frac{3}{5} \frac{GM_{DM}}{R_{DM}} \left[1 + \frac{1}{x}
\left(\frac{5}{2} - \frac{3}{2} \frac{1}{y^2} \right) \right]
\label{sigma}
\end{eqnarray}

\noindent
The factor 3/5 in front to each term stems from the politropic description of the potential energy
$E_G = \frac{3}{5-n}\frac{G M^2}{R}$  where $n$ is the politropic index, which for an homogeneous
distribution of mass is $n=0$.
The relations given by eqs. (\ref{sigma}) represent the new virial conditions for an ideal composite
galaxy made of DM and BM  with concentric spherical and homogeneous distributions.
These virial conditions are fully equivalent to the one of eq.\ref{eq1} and would immediately
generate the relation of eq. \ref{eq2}.
The weakest issue of the above formalism is the assumption of homogeneity of the two subsystems.
Nevertheless, for the use we are going to make of the above equations, this will be almost irrelevant.

To proceed further we need to know the fraction of BM originally in form of gas that is actually converted to stars.
Numerical simulations of ETGs formation indicate that a large amount of gas is left over by the star formation
activity and is heated up by feedback effects escaping in some cases the potential well.
The typical stars to gas ratio is $\sim 0.25$
\citep[]{ChiosiCarraro2002,Chiosietal2012,Chiosi2014,Merlin2006,Merlin2007,Merlin2010,Merlin2012}.

Now we consider only the equations for the BM component, but limited to the stars:

\begin{eqnarray}
2 (E_{s})_{k} &=& \frac{3}{5} \frac{GM_{s}^2}{R_{s}} + \frac{3}{5} \frac{GM_{s}^2}{R_{s}} \frac{x}{y^3}\\
\sigma_{s}^2 &=& \frac{3}{5} \frac{GM_{s}}{R_{s}}  \left(1 + \frac{x}{y^3} \right).
\end{eqnarray}

At this point, to arrive at the $L-\sigma$ relation we follow a method different from the one used for eq.
\ref{eq2}.  We look for the  relationship between $M^*$ and  $R_s$.
To this aim, we follow the formulation of the \MR\ relation developed by \citet{Fanetal2010}.
Assuming spherical symmetry for the sake of simplicity and the standard ratio $M_{DM}/M_{BM} \simeq 6.6 \equiv
x$, the mass-radius relation for proto-galaxies made of DM and BM with total mass $M=M_{DM}+M_{BM} \simeq
1.15 \times M_{DM} \simeq M_{DM} $ for all practical purposes, is given by

\begin{equation}
\frac{4\pi}{3} R_{DM}^3 =  \frac{M_{DM}}{\lambda \rho_{u}(z) }
\label{prone}
\end{equation}

\noindent
where $\rho(z) \propto (1+z)^3$ is the density of the Universe at redshift $z$ and $\lambda$ the factor for
the density contrast of the DM halo. This expression is of general validity whereas the function $\lambda$
depends on the cosmological model of the Universe (including the $\Lambda$-CDM case). All details can be
found in \citet[][their eq. 6]{Bryan_Norman_1998}.

In the context of $\Lambda$-CDM cosmology, \citet{Fanetal2010} have adapted eq. \ref{prone} to provide an
relationship between $R_s$ and $M^*$. They assume that over the Hubble time each halo  that collapsed at
redshift $z_f$  generate a
stellar  mass $M^*$. The stellar mass $M^*$ is then expressed by the ratio
$M^* = M_{DM}/ \theta_s$ where $\theta_s$ is taken from numerical simulations of galaxy formation.
Finally,  the  half-mass radius is $R_s$ is:

\begin{equation}
R_{s}=0.9 \frac{S_S(n)}{0.34}\frac{25}{\theta_s} \left( \frac{1.5}{f_s} \right)^2
\left( \frac{M_{DM}}{10^{12} M_\odot} \right)^{1/3} \frac{4}{(1+z_{f})}.
\label{mrr}
\end{equation}
where $R_s$ is in kpc.
\noindent
As already anticipated, on average the efficiency of star formation is such that only a fraction of the
original gas is converted to stars
(typically a fourth of it). Let us call this fraction $Q_s$ and consider it as an adjustable parameter,
i.e.  $Q_s M_{BM}= M^*$. This quantity can be soon related to the  ratio $x$ of
\citet{Caimmi2003,Caimmi2007,Caimmi2009} and the parameter $\theta_s$ by  $\theta_s =  x/Q_s$. Another
useful relation is $\theta_s = (M_{DM}/M_{BM} )/Q_s$, that for the assumed
cosmology and efficiency of star formation gives $\theta_s \simeq 25$.
The quantity $S_S(n_S)$ is a coefficient related to the S\'ersic indexes $n_S$ and to the ansatz
$R_{s}=S_S(n_S)R_g$ relating gravitational and stellar mass radii, $f_{s}$ the velocity dispersion of the
stellar component with respect to that of DM.  All these quantities have been evaluated by
\citet[][to whom we refer for all the details]{Fanetal2010}: $S_S(n_S)=0.34$, $f_{s}=1$. Also in this
case the exact values for all these quantities  are not mandatory here. It is worth noting that
Eq. \ref{mrr} links the stellar half-mass radius $R_s$ to the mass $M_{DM}$ of its DM halo host.

We define now the three constants $K_1$, $K_2$ and $K_\sigma$:
\begin{eqnarray}
K_{1} &=& 0.9 \frac{S_S(n)}{0.34}\frac{25}{\theta_s} \left( \frac{1.5}{f_s} \right)^2 \theta_s^{1/3}  \\
     K_2 &=& 4 K_1 \left(\frac{1}{10^{12} M_\odot}\right)^{1/3}   \\
         K_\sigma &=& \left[\frac{3}{5} \frac{G}{K_2 Q_s} \right]
\end{eqnarray}
\noindent
The half-mass radius $R_s$ of Eq.(\ref{mrr}) can be recast as:

\begin{equation}
R_{s} = K_{2} \times {M^*}^{1/3} \left(\frac{1}{1+z_f}\right),
\label{mrr_2}
\end{equation}
and the velocity dispersion  of the stars $\sigma_s$  can be written:
\begin{equation}
\sigma = K_{\sigma} M^{*{1/3}} \left( 1 + \frac{x}{y^3} \right)^{1/2} (1+z_f)^{1/2},
\label{sig_mass_zeta}
\end{equation}
If we write $M^*$ as $\Gamma L$, where $\Gamma$ is the  stellar mean mass-to-light ratio, in logarithmic
variables  we obtain:
\begin{eqnarray}
\log(L) & = & 3\log(\sigma)-\log(\Gamma)-3\log(K_{\sigma})+ \nonumber \\
\label{L_sig_zeta}
&   & - \frac{3}{2}\log(\left( 1 + \frac{x}{y^3} \right)-\frac{3}{2}\log(1+z_f) + const.
\end{eqnarray}
Eq. \ref{L_sig_zeta} has eventually taken  a form mimicking the FJ relation. Here we see that the
exponent of the $\sigma$-term   is now equal to 3 (instead of 2) that mirrors the slope of the assumed
mass-radius
relation of eq. \ref{mrr}, there is term containing the parameters x and y related to the presence of dark
and baryonic matter, the term related to the galaxy formation redshift,  and a final $const$ that is related
and fixed  by the units adopted for the different quantities in usage.

Finally, we point out  that in the case of $M_{DM} \simeq  0$ (no Dark Matter) the parameter $x
\rightarrow 0 $,  eq. \ref{eq2} cannot be recovered beacause of the different power for the velocity
dispersion and the term $(1+z_f)$ which is not
related to the presence of DM but to the starting hypothesis  of the  proto-galaxy collapsing at at redshift
$z_f$. If we drop it or simply do not make it explicit the formal recover of eq. (\ref{eq2}) is
straightforward.

In conclusion, the virial theorem of eq. (\ref{eq1}) and the \Lsig relation of \ eq. (\ref{eq2})
should contain  additional terms, and  the general \Lsig\ relation should be given eq. (\ref{L_sig_zeta}).

\section{Model galaxies: bursts of star formation and Mergers}\label{Appendix_B}

In this section we shortly present the Monte-Carlo method we ave used to describe bursts of star
formation in galaxies of any mass, age and mean metallicity and mergers among galaxies of different mass,
age and mean metallicity.

As a first step, we approximate the complex mix of stellar populations inside a galaxy of mass
$M$ with a SSP of suitable age $T_G$,  metallicity $Z_G$ and the same mass.  $T_G$, $Z_G$  are the age and
metallicity reached at the peak of star formation, which according to current galaxy models occur shortly
after the formation of the
galaxy itself
\citep[][]{Chiosi_etal_2017, Sciarratta_etal_2019} so that $T_G$ ($z_G$) roughly corresponds to the formation
time (redshift).   With $T_U$  the present age of the Universe
for the adopted cosmological scenario, then
$T_{G,z}=T_U - T_G$ is the age (redshift) of the Universe when the galaxy was born.

In the following, we will present two paradigmatic cases:

(i) An already in place galaxy via the
initial major episode of star formation, which later undergoes an additional episode of star formation of
minor intensity (thereafter referred  as \textit{burst} case). With simulations of this kind, we explore the
consequences of adding young stellar components to already evolved stellar assemblies,
in other words we can estimate the effect of a rejuvenation event on an otherwise passively
evolving stellar system.
This is the analog of simulating either completely wet mergers or a minor stellar activity
for any internal
reason (eg. re-use of the gas shed by RGB and AGB stars).

(ii)  The other interesting case to consider
is  the case of a \textit{merger} of two galaxies of different mass, age, and metallicity. This
 would simply tell us how the photometric properties of each of the two subsystems added together would give rise to a
 new photometric appearance of the composed system even in absence of companion star formation.
This is the analog of a random combination of wet and dry mergers.

\textsf{Bursts}.
The  age $T_G$  of the initial star forming episode is supposed
to fall in the age range $T_{G,max} > T_G > T_{G,min}$.
Subsequently,  a burst of star formation engaging a certain percentage of the mass (typically up to about
$10\%$) is supposed to occur at any age $T_B$ comprised between $T_{B,max} = T_{G,min}$ and  the present
time (more precisely
$T_{B,min}=0.1$ Gyr,  the minimum age in the SSP grids).

The rest of the procedure is quite
simple: first, we  take  the fluxes from SSPs of different metallicities,
normalize them to unit of mass (with the Salpeter IMF and $M_{l}=0.1\, M_\odot$
and $M_{u}=100\, M_\odot$, $M_{SSP}=5.826\, M_\odot$), and then multiply them
by the mass of the galaxy. Next, we randomize ages and masses of the seed SSP-galaxies together
with the age and mass percentages of the burst episode. To this aim, it is more convenient to express
the age and masses in
terms of their logarithms, in order to avoid non-uniform distribution in the randomly chosen values.
The ages (written with lower case symbols to remind the reader that they are expressed as logarithms)
of the seed galaxies are given by

\begin{equation}
t_{G} = t_{G,max} - r\,(t_{G,max} - t_{G,min})
\label{eqageseed}
\end{equation}

\noindent and those of the bursts by
\begin{equation}
t_{B} = t_{B,max} - r\,(t_{B,max} - t_{B,min})\,;
\label{eqageburst}
\end{equation}

\noindent $r\in(0,1)$ is a random, always different number. Similar procedure is made for the mass
of the seed SSP-galaxy,
which spans the range $10^7$ to $10^{12}$ $M_\odot$, and the mass percentage $p_B$ of the burst mass with respect to
the mass of the host galaxy. The percentage $p_B$ goes from 0 to 0.5. Therefore,
the relative contribution of the two components to the total flux (magnitudes in any pass-band) is
given by

$$f  = (1 - p_B)\,f_G   + p_B \,f_B$$

\noindent with obvious meaning of the symbols.
Finally, since the SSP fluxes (magnitudes and colors) depend on both age and metallicity
and we know that this latter in turn increases with the galaxy mass, we have taken this into
 account by adopting an empirical
mass-metallicity relation that is based on chemical models of galaxies and observational data and is
presented here in Table \ref{mass_metals} taken from \citet{Sciarratta_etal_2019}. Shortly speaking,
metallicity is for simplicity binned in terms of logarithmic mass. Finally, for each burst, the metallicity is
for simplicity chosen
to be equal to that of the seed galaxies. This means that, at variance with
mergers, metallicities will not mix together.

\textsf{Mergers}.
The mass, metallicity and age of each galaxy  are derived using the same procedure as above, the only
difference being that the permitted age interval extends now  over nearly the whole Hubble time i.e.
$T_{G,max} - T_{G,min} \simeq T_{G,max}$.
Denoting with $T_{G,j}$ the  age of the $j-$th component of the merger
(a single event for simplicity) and using the
logarithmic notation, the ages  $t_{G,j}$ are

\begin{equation}
t_{G,j} = t_{G,max,j} - r\,t_{G,max,j} \label{eqageseedmerg}
\end{equation}

\noindent where $r$ is the random number.

When  two galaxies merge together, in the resulting
bigger object there are stars from both initial components. At each time, their contribution to the total flux
is first suitably
shifted by the age difference between the two components to set up a common clock
and then weighed by the mass of each component.
To keep our simulations simple,  mergers occur between of single pairs of galaxies and
 the case of multiple mergers is not considered. Furthermore, we will use only SSPs with solar partition of
$\alpha-$elements, i.e $[\alpha/Fe]=0$.

\begin{table}
\begin{center}
\caption{ Empirical mass-metallicity relation for SSP-galaxies. Logarithmic masses are in solar units.}
\label{mass_metals}
\begin{tabular}{|c| c| c| c| c| c| c| }
\hline
$\log M$   & 7 - 8  & 8 - 9 & 9 - 10 & 10 - 11 & 11 -12  & 12 - 13 \\
$Z_{min}$ & 0.0004 & 0.001 & 0.004  & 0.008   & 0.019   & 0.040   \\
$Z_{max}$ & 0.0010 & 0.004 & 0.008  & 0.019   & 0.040   & 0.070   \\
\hline
\end{tabular}
\end{center}
\end{table}
 \end{appendix}

\end{document}